\documentclass[12pt]{article}

\usepackage{color,epsf}
\usepackage{graphicx}
\usepackage{amsmath,amssymb}

\newcommand{\insertfig}[2]{\mbox{\epsfxsize=#1cm \epsfbox{#2.eps}}}

\font\cmss=cmss12 
\def\1{\hbox{{1}\kern-.25em\hbox{l}}}
\def\bfZ{\relax{\hbox{\cmss Z\kern-.4em Z}}}

\begin{document}

\begin{titlepage}

\vspace*{2ex}

\begin{center}
\Large \bf
Sum rules and dualities for generalized parton distributions:
is there a holographic principle?
\end{center}

\vspace{15mm}

\centerline{\bf   K.~Kumeri{\v c}ki$^{a}$, D.~M\"uller$^{b}$,
K.~Passek-Kumeri{\v c}ki$^{c}$
}

\vspace{8mm} \centerline{\it $^a$Department of Physics, Faculty of
Science, University of Zagreb} \centerline{\it P.O.B. 331,
HR-10002 Zagreb, Croatia}

\vspace{5mm} \centerline{\it $^b$Institut f\"ur Theoretische
Physik II, Ruhr-Universit\"at Bochum} \centerline{\it D-44780
Bochum, Germany}

\vspace{5mm} \centerline{\it $^c$Theoretical Physics Division,
Rudjer Bo{\v s}kovi{\'c} Institute} \centerline{\it P.O.Box 180,
HR-10002 Zagreb, Croatia}

\vspace{15mm}

\centerline{\bf Abstract}

\vspace{0.5cm}

\noindent
To leading order approximation, the physical content of
generalized parton distributions (GPDs) that is accessible in
deep virtual electroproduction of photons or mesons is contained
in their value on the cross-over trajectory. This trajectory
separates the $t$-channel and $s$-channel dominated GPD regions.
The underlying Lorentz covariance implies a correspondence between
these two regions through their relation to GPDs on the cross-over
trajectory. This point of view leads to a family of GPD sum rules
which are a quark analogue of finite energy sum rules and it
guides us to a new phenomenological GPD concept. As an example, we
discuss the constraints from the JLab/Hall A data on the dominant
$u$-quark GPD $H$. The question arises whether GPDs are governed
by some kind of holographic principle.

\vfill


\end{titlepage}

\tableofcontents

\newpage

\section{Introduction}

Generalized parton distributions (GPDs)
\cite{MueRobGeyDitHor94,Rad96,Ji96a} are partonic  amplitudes
which contain the  non-perturbative physics of perturbatively
factorized exclusive leptoproduction amplitudes
\cite{ColFraStr96,ColFre98}. Depending on the parton momentum
fractions, a GPD is interpreted either as the probability
amplitude of meson-like $t$-channel exchange or as 
$s$-channel exchange of a parton. Compared to parton distribution
functions (PDFs) and distribution amplitudes, they reveal the
partonic content of hadrons, protons in particular, from a
complementary perspective; see for reviews \cite{Die03a,BelRad05}. For
instance, the partonic angular momentum, naturally defined by the
decomposition of the angular momentum tensor in terms of quark and
gluonic degrees of freedom, is given by the first Mellin moment of
certain GPDs \cite{Ji96}. Quantification of such partonic
decomposition of the proton spin is one of the main goals for
measurements of GPD related processes.

During the previous decade, experimental effort has been devoted to
the measurement of deep virtual Compton scattering (DVCS)
\cite{Adletal01,Airetal01,Steetal01,Chekanov:2003ya,Aktas:2005ty,Cametal06,Airetal06,Cheetal06,Mazetal07,Aaretal07,Giretal07,Airetal08}
and production of mesons, see e.g.
Refs.~\cite{Breetal98,Adletal99,Breetal99,Airetal07,GuiMor07},
yielding an increasing amount and precision of experimental data.
On the other hand, GPDs are intricate functions, depending on the
parton's longitudinal momentum fraction ($x$), the skewness
($\eta$), the momentum transfer squared ($t$), and the
factorization scale ($\mu^2$ --- usually set equal to the virtuality ${\cal Q}^2$
of the probe). This makes their extraction from experimental
measurements, where they enter as a convolution with a
hard-scattering amplitude, a rather awkward task. Since the
partonic momentum fraction dependence of GPDs is integrated out in
the amplitude, the degrees of freedom that can be constrained by
experiment do not provide the momentum fraction shape of GPDs for
fixed $\eta$. Consequently, GPD moments, including the angular
momentum of quarks, cannot be directly revealed.

The analysis of deep virtual exclusive processes is  based on
GPD ans\"atze and models which take into account that GPDs
reduce by sum rules to form factors and in the forward limit to
PDFs. As the basic GPD
ansatz, written in essence as a product of form factor and PDF,
the popular VGG code \cite{VanGuiGui98} successfully describes the
first DVCS beam spin asymmetry measurements at HERMES
\cite{Airetal01} and CLAS \cite{Steetal01}, but it fails in the case
of more recent data on the unpolarized DVCS cross section,
measured by the Hall A collaboration \cite{Cametal06}. Other GPD
models \cite{BelMueKir01,KirMue03,GuzTec06,PolVan08} are similarly
(un)successful in describing DVCS observables measured in fixed
target experiments. In collider kinematics, no published GPD
model, which is (a) mathematically consistent, (b) has a
reasonable $t$-dependence, and (c) satisfies the evolution
equation, is able to describe the DVCS data at leading order (LO)
accuracy. Taking into account radiative corrections, the measured
DVCS cross section can be described, as advocated in
Refs.~\cite{FreMcD01a,FreMcD01c}, and fitted in
Refs.~\cite{KumMuePasSch06,KumMuePas07}. Also, Regge inspired
modelling of off-shell amplitudes \cite{Lag00} can be tuned to be
consistent with experimental data
\cite{CanLag02,CapFazFioJenPac06}. If unitarity constraints are
taken into account in the fixed target kinematics, the model
\cite{CanLag02} is in fair agreement with the unpolarized DVCS
cross section measurement \cite{Lag07}. Thereby, it provides a
physical interpretation in terms of hadronic degrees of freedom,
where the desired view inside the nucleon cannot be delivered.

Thus, although GPD modelling is guided by the relation of GPDs to PDFs
and form factors, by Regge phenomenology, various model
calculations, lattice simulations etc., the resulting models are only
partially able to describe experimental data. This, in combination
with the wide-spread implicit belief that the resulting GPD models
might be realistic in a more general way, leads then to endless
discussions, such as 'Which model is right, A or B?' (when both
are maybe wrong) or `GPDs or not GPDs?', to claims that GPDs can
be constrained from present lattice data, and to `a conjectured
proof'  of the breakdown of the  GPD formalism (for some details
see \cite{KumMuePas07a}). Sometimes, criticism of GPD
\emph{models} is unjustly extended to GPD \emph{representations}.
In our opinion it is unclear whether sufficient understanding of
GPDs is reached in the proposed models. Below we shall demonstrate
that there should be {\em no problem} to describe presently
measured DVCS observables within the GPD formalism. Hence, we should
question the present phenomenological  approach,
which is based on GPD models that we consider to be  \emph{ad hoc}
constructions.

We would like to come back to two basic questions, which should be
clearly answered.
\begin{itemize}
\item
What GPD information can be revealed from experimental data?
\item
What do we learn from this information?
\end{itemize}

\noindent Both above questions have been addressed in the momentum
fraction representation and, more recently, in the framework of
the `dual' GPD parameterization \cite{Pol07,Pol07a,PolVan08}. We
shall recall the known partial answers to those questions in
the momentum fraction representation and add a rarely recognized GPD
aspect, namely, duality \cite{MueSch05,KumMuePas07}, with the
focus on the phenomenological applicability. Before we give the
technicalities in the main body of the paper, we briefly outline
the answers here. In our opinion it does not matter in which
representation they are formulated, and in this paper  we prefer the
momentum fraction representation.

The answer to the first question arises from analyticity and it
has been clearly spelled out for a fixed resolution scale in
Ref.~\cite{Ter05}.  Let us recall that the DVCS amplitude is
considered to be a holomorphic function, see, e.g.,
Refs.~\cite{FraFreGuzStr97,Che97,Ter05,KumMuePas07,DieIva07,AniTer07},
whose real part can be expressed in terms of its imaginary part,
i.e., its $s$- and $u$-channel discontinuities, via a single
variable dispersion relation for fixed $t$. Moreover, to LO the
imaginary part is given by the GPD on its \emph{cross-over
trajectory} ($\eta=x$), which separates the meson-like $t$-channel
exchange interpretation of the \emph{central region} ($ |x| \le
\eta$) from the partonic $s$-channel exchange view of the
\emph{outer region} ($ \eta \le x \le 1$). Hence, apart from the
cross-over trajectory the GPD for fixed resolution scale cannot be
experimentally accessed. The dispersion relation gives us a
considerable handle on the GPD at the cross-over trajectory and
constrains its functional form, even if the modulus and  phase of
the amplitude are measured only in a restricted phase space.
Therefore, the answer to the first question  guides us to a simple
phenomenological concept in which the dispersion relation is used
to pin down the value of the GPD on the cross-over trajectory.

Suppose we know now the GPD at its cross-over trajectory from an
(ideal) measurement for fixed photon virtuality ${\cal Q}^2$, as
illustrated below in Fig.~\ref{Fig-GPDans}. The deconvolution
problem, which then pops up, is how the GPD at $\eta=x$ is related
to its value at any $\eta$, i.e., for the whole support.  Lorentz
covariance guarantees that a GPD possesses a duality property
\cite{MueSch05,KumMuePas07}, i.e., it can be mapped from the outer
region  into its central one, where the GPD is continuous on the
cross-over trajectory. Obviously, one can draw a priory  any
surface over the outer region of the GPD support that connects the
curves given by the GPD at $\eta=x$ and $\eta=0$ for $0 \le x\le
1$, see Fig.~\ref{Fig-GPDans}. Positivity constraints, see
Ref.~\cite{Pob02,Pob02a} and references therein, lead to constraints on
these surfaces. The dual surfaces in the central region are then
governed by Lorentz covariance. Each of these surfaces corresponds
to a GPD model, and so the deconvolution
problem has infinite solutions.
The problem with present ad hoc GPD models is that usually
one solution is picked out, with parameterization of the GPD at the
cross-over trajectory being too rigid.
We stress that evolution
of GPD with change of the resolution scale provides a supplementary
handle on the deconvolution problem, as advocated in
Ref.~\cite{Fre99}. However, for a small lever arm in the photon
virtuality,  only if some holographic principle exists, rigidly
relating shape of this surface to values along the $\eta=x$
cross-over trajectory, the deconvolution problem can be strictly solved.  But
even if this hypothetical principle remains unknown, one still has
various theoretical and phenomenological tools at hand to address the
deconvolution problem and, in particular, the related problem of
decomposition of the quark angular momentum.

In this paper we introduce a  new set of such tools, namely GPD
sum rules, and we  illustrate how an alternative phenomenological
approach to DVCS might look like. The outline is as follows. To
shed light on unrealistic features of the GPD models, presently used
in phenomenology, we give in Sect.~\ref{Sec-GPD-mod}  a short
overview of experimental findings and their confrontation with GPD
models, mainly for the DVCS process. The reader who is not
interested in this phenomenological aspect can safely skip this
section. In Sect.~\ref{Sec-DuaSumRul} we employ previous results
\cite{MueSch05,KumMuePas07} to clearly spell out that GPDs have an
internal duality, relating the outer and central regions. We
derive then various GPD sum rules. In Sect.~\ref{Sec-Pha} we
discuss the phenomenological applications of these sum rules  and
also give a preliminary example for fixed target kinematics.
Finally, in Sect.~\ref{Sec-Out} we summarize our findings and conjecture that
GPDs should be governed by a holographic principle,  which might
be naturally related both to the duality interpretation of strong
interaction phenomena, reviewed, e.g., in Ref.~\cite{Col77}, and
to the AdS/CFT conjecture \cite{Mal97,GubKlePol98,Wit98}.

\section{Mainstream GPD models versus experiment}
\label{Sec-GPD-mod}

Numerous experimental, theoretical, and phenomenological efforts,
reviewed in Refs.~\cite{Die03a,BelRad05}, have been spent during
the first decade of GPD phenomenology in an attempt to go
beyond the basic GPD ansatz, given by the product of PDF and
elastic form factor. Fortunately, there is tremendous progress on
the experimental side and, certainly, some theoretical
understanding is reached, too. Unfortunately, some rather
hypothetical elements became an important part of the mainstream
phenomenology, while at the same time some well-understood
theoretical features are incompletely implemented. In what
follows, we would like to make it clear that the observed
phenomenological problem, namely, that certain GPD models fail
to describe the DVCS data \cite{Cametal06,GuiMor07,Gui08,PolVan08},
indicates only that the ad hoc GPD model approach in its present
form  is an inefficient tool for the future and that any general
conclusion about the GPD formalism drawn from the failure of
certain GPD models should be considered as a speculation.  Below we briefly describe the
situation and attempt to separate different problems, which are
often mixed up.

\subsection{GPD basics and terminology}

GPD models, used in phenomenology so far, are mainly based either on a
spectral representation for the Green functions of light-ray operators,
so-called double distributions (DDs)
\cite{MueRobGeyDitHor94,Rad97}, or on the collinear conformal partial
wave expansion%
\footnote{The group theoretical aspects and QCD applications of
the collinear conformal group are given in
Ref.~\cite{BraKorMue03}.}
\cite{BelGeyMueSch97}, known in various versions
\cite{Shu99,Nor00,BalBra89,KivMan99b,MueSch05,ManKirSch05,KirManSch05a}.
Since they all by definition represent the same
field theoretical object, any GPD model, as
long as it respects basic properties, such as Lorentz covariance, can
be represented in any of the named representations. Another popular
GPD representation is given as overlap of light-cone wave functions
\cite{DieFelJakKro98,BroDieHwa00,DieFelJakKro00}.
We emphasize that
the one-to-one correspondence of the DD and
overlap light-cone wave function representation
\cite{DieFelJakKro98,BroDieHwa00,DieFelJakKro00} should hold on
general field theoretical grounds, too. This has been demonstrated
within wave function models in
Refs.~\cite{MukMusPauRad02,TibDetMil04,HwaMue07}. In the overlap
light-cone wave function representation the positivity constraints
\cite{Pob02,Pob02a} are explicitly satisfied.%
\footnote{
The solution of these general positivity constraints
can be interpreted as a two body light-cone wave function overlap
representation for a collective spectator \cite{Pob03}.
} 
These constraints are exactly valid at LO approximation;
however, their power drastically diminishes if one calculates the
physical amplitude.

\subsubsection{Double distribution representation}

The DD representation arises directly from diagrammatical considerations
\cite{MueRobGeyDitHor94,Rad97}. We
consider it to be the most general representation that inherits
properties of field theory. Here,
the partonic quantum numbers are the momentum
fractions of two partons. We recall that the first  sophisticated GPD ansatz was
elaborated in Refs.~\cite{Rad98a,Rad98} for $t=0$ via a  DD
ansatz, which factorizes in a PDF and a {\em single}
variable profile function whose shape, viewed as a meson
distribution amplitude, was assumed to be convex. For brevity,
we call this the Radyushkin's DD ansatz (RDDA).
Because of its convenient parameterization, it was adopted by many
model builders in various $t$-decorated versions and within
generalizations of the profile function; however, the convexity of
its shape was not touched upon. In particular, the RDDA \cite{Rad98a,Rad98}
has been utilized in Ref.~\cite{GoePolVan01} to build up a GPD
model, guided by the results of the chiral quark soliton model
($\chi$QSM), decorated with  a factorized or Regge inspired
$t$-dependence, and based on certain PDF parameterizations. This
GPD model is now numerically implemented in the popular VGG code
\cite{VanGuiGui98}.   Here a descendant of the $D$-term has been
utilized to parameterize the GPD $E$ in terms of the quark
angular momentum.  We follow the common terminology and call this
code the VGG model. We note that VGG results might be different from
other GPD models that were written down in a similar fashion
within RDDA, e.g., in Ref.~\cite{BelMueKir01}.

It was already emphasized  in Refs.~\cite{Rad98a,Rad98,Rad98c} that more
realistic GPD models  should possess an intricate interplay of
momentum fraction  and momentum transfer squared dependence. It
turned out that guidance for a more realistic $t$-dependence comes
from field theory inspired models. Such an improvement is now
implemented in  GPD models for the pion \cite{MukMusPauRad02} and
nucleon \cite{TibDetMil04,HwaMue07}.
All these investigations started from
overlap representations with power-law wave functions, and, aiming
at different goals, they finally provided rather analogous
expressions for the resulting DDs.%
\footnote{It was
observed in Ref.~\cite{MukMusPauRad02} that the functional form of
light-cone wave functions cannot be arbitrarily chosen to derive
the DD representation. Therefore, the authors of
Ref.~\cite{TibDetMil04} started with a covariant quark model to
obtain GPD models, formulated in DD
representation. In Ref.~\cite{HwaMue07} it was stated that in the
light-cone wave function overlap representation the longitudinal
and transversal degrees of freedom are tied to each other due to
Lorentz symmetry. Hence, arbitrary modelling of light-cone
wave functions will usually violate the polynomiality property of
GPD moments, see, e.g., Ref.~\cite{BofPas07}.
}
Such models can also be obtained from the evaluation of (general) triangle
Feynman diagrams. To emphasize the physical content of these GPD
models, one might also denote them `collective spectator quark
models'.%
\footnote{There is a huge literature in which GPDs have been
studied within dynamical model approaches. The reader may find
references in Refs.~\cite{Die03a,HwaMue07,BofPas07}.
}
For both $\eta=0$ and $\eta=x$ the $t$-dependence of these
models dies out in the limit $x\to 1$; see, e.g., discussion in
Ref.~\cite{HwaMue07}. The behavior in the former case coincides
with lattice measurements of generalized form factors
\cite{Hagetal03,Hagetal04,Hagetal04a,Gocetal03,Gocetal05,Edwetal06,Hagetal07}.
A more general DD ansatz in the form of a
factorized RDDA with the common improvement of Regge behavior
has been written down for certain proton GPDs in
Ref.~\cite{HwaMue07}; for the pion case see
Ref.~\cite{MukMusPauRad02}.  We have also convinced ourselves that
one should not insist on a convex shape of the profile function, and
could, e.g., add 'higher' modes, to get rather flexible GPD
models.
In this way one can replace ad hoc GPD model approach with a
more flexible fitting procedure.

\subsubsection{Conformal partial wave expansions}

The expansion of the GPD in collinear  $SL(2,\mathbb R)$
conformal partial waves arises naturally from the solution of the
evolution equation to LO accuracy. The parameterizations, applied in phenomenology,
are named
`dual' \cite{PolShu02} or Mellin-Barnes representation
\cite{MueSch05}. In our understanding, the name `dual'
primarily refers to the use of crossing
symmetry \cite{MueRobGeyDitHor94,DieGouPirTer98,Ter01}, which is, by
itself, independent of the chosen representation.  For the sake of
clarity, we follow the authors of Ref.~\cite{PolShu02} and
denote their representation `dual' in quotation
marks. The term dual (without quotation marks) we use for a
general GPD property, as used above and explained further in
Sect.~\ref{Sec-DuaSumRul}. This concept was partially utilized for
evolution kernels \cite{GeyDitHorMueRob88,MueRobGeyDitHor94}  and
has been  adopted to GPDs \cite{MueSch05}, supplemented by a
constructive proof \cite{KumMuePas07}. As emphasized in
Ref.~\cite{MueSch05} it is also tied to crossing, again
independently of the representation. The term Mellin-Barnes
representation refers to a specific integral representation of the
collinear conformal partial wave expansion, closely related to the
well-known  Sommerfeld-Watson transform, in which internal GPD
duality is manifestly incorporated,
cf.~Ref.~\cite{ManKirSch05,KirManSch05a}.

In the Mellin-Barnes
representation, besides the complex-valued conformal spin, the
angular momentum of the $t$-channel SO(3) partial wave
might be employed as a label. In the `dual'
parameterization a forward-like momentum fraction (conjugated
variable to the conformal spin) and the difference of integral
conformal spin and $t$-channel angular momentum  appear as
labels. This makes contact to hadronic physics \cite{PolShu02}, in
particular to Regge phenomenology \cite{KumMuePas07}.  In the
minimalist version of this model the expansion of conformal
moments in terms of SO(3) $t$-channel partial waves is restricted
to the leading one. The inclusion of the next-leading SO(3)
partial wave has been called minimal `dual' parameterization
\cite{PolShu02}. More recently it has become clear how the
inclusion of non-leading  SO(3) partial waves can be effectively
achieved in the `dual' parameterization \cite{Pol07a}. The
minimalist and minimal version in the `dual' parameterization
corresponds to the  leading and next-leading SO(3) partial wave
approximation.  We use the terminology: leading (i.e., minimalist)
SO(3) partial wave and next-leading (i.e., minimal) SO(3) partial
wave approximation \cite{KumMuePas07}. GPD modelling in the `dual'
and the Mellin-Barnes integral parameterizations is presently done
with different phenomenological emphases. In the former
parameterization the model aspect with respect to the
approximation of SO(3) partial waves, see Sect.~\ref{Sec-Pha-Str},
is pronounced \cite{Pol07,Pol07a}, while in the latter, up to now
applied only to small $x_{\rm Bj}$ physics, the Regge aspect
\cite{KumMuePas07} and the resummation of SO(3) partial waves
\cite{KumMuePasSch08} is given priority.

\subsubsection{Other GPD expansions}

One might also introduce an expansion with respect to a complete
basis of orthogonal polynomials, such as Gegenbauer
\cite{BelGeyMueSch97} or  Bernstein \cite{AhmHonLiuTan07}.%
\footnote{Interestingly, in Ref.~\cite{AhmHonLiuTan06}
the authors started from a covariantly formulated spectator
model; however, consider Lorentz covariance (polynomiality of GPD
moments) as a DD hypothesis. Finally, they are
doing  GPD modelling for $\eta=0$. To include skewness dependence,
they parameterize the GPD in terms of a few Bernstein polynomials
and rely on lattice data \cite{AhmHonLiuTan07}.
}
Such  expansions can in principle be expressed in the two
considered ones. The expansion in terms of Bernstein polynomials
has been proposed once in deep-inelastic scattering (DIS) \cite{Ynd78}; 
however, it could not compete with the
standard approaches, employed in global DIS fits. Also, we know
from experience that the expansion of GPDs with respect to a
complete basis of orthogonal polynomials turns out to be
numerically inefficient at smaller values of $x_{\rm Bj}$
\cite{BelMueNieSch99}.

\begin{figure}[t]
\centerline{\includegraphics{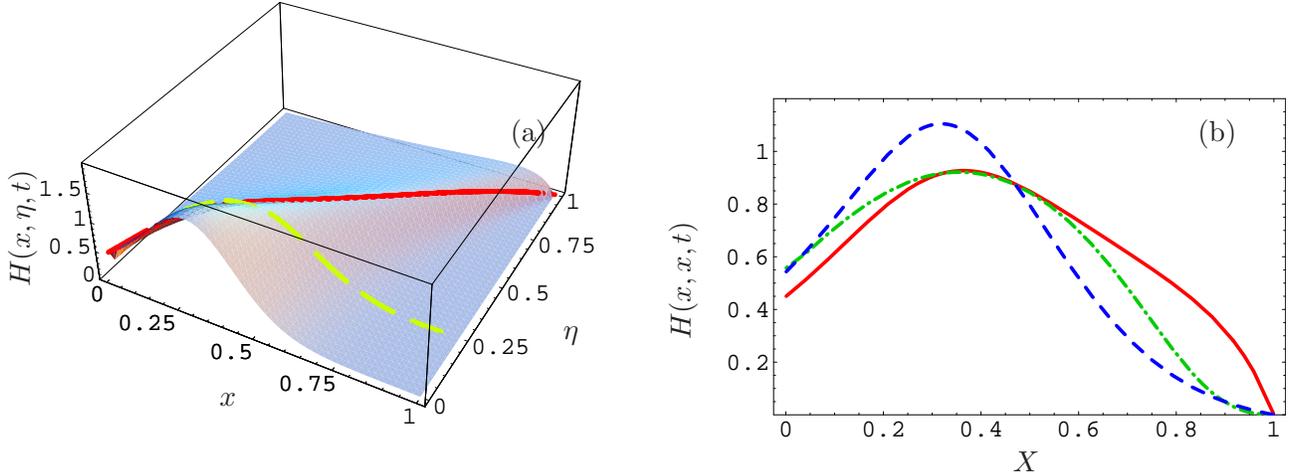}}
\caption{\label{Fig-GPDans}
The left panel displays the spectator  model of
Ref.~\cite{HwaMue07}, as specified there in Eq.~(25), with
$t=t_{\rm min}-0.25\,{\rm GeV}^2$. The red solid curve visualizes
the experimentally accessible information for fixed photon
virtuality, i.e., the GPD at the cross-over trajectory $\eta=x$,
while the dashed one is an example curve that can be used to evaluate
the real part of CFFs via Eq.~(\ref{Def-AmpLO}). In the right
panel the same GPD at $\eta=x$ (solid)  is displayed versus the
Bjorken-like  variable $X=2x/(1+x)$ and it is compared with a
VGG-like (dashed) and the leading SO(3) partial wave approximation
of the Mellin-Barnes (dash-dotted) ansatz. Both ans\"atze   are
based on the zero skewness spectator GPD  model and possess a
factorized $t$-dependence.
}
\end{figure}

\subsection{Phenomenological applications}

The phenomenological application of GPDs for the deep virtual
leptoproduction of mesons or photon is based on factorization
theorems, proven to leading order in the $1/{\cal Q}$ expansion
\cite{ColFraStr96,ColFre98}. Thereby it is assumed that GPDs are
continuous at the cross-over point. It is commonly assumed that
the GPD framework is quantitatively applicable for DVCS at a scale
of the order of one or few $\mbox{GeV}^2$, while for meson
production it might be of the order of 10 $\mbox{GeV}^2$.
Unfortunately, only very little is known about the size of power
suppressed contributions and it remains an open problem to pin
down the onset scale for the GPD formalism. This problem might
also be addressed from the phenomenological side. In the following we
assume that the GPD formalism is quantitatively applicable to
leading order of the perturbative expansion in the QCD coupling
constant for DVCS measurements in JLAB, HERMES and H1/ZEUS
kinematics, where the photon virtuality is larger than 1
$\mbox{GeV}^2$.

The specific GPD models are all more or less constrained by the
same phenomenological input, PDFs, form factors, and sometimes
model or lattice estimates. Roughly spoken, the difference in the
physical content of various GPD models at $\eta=x$ is more or less
accidental and arises from simplifications, mainly done for
convenience, in a given representation. Obviously, the model
dependence is still visible in the described deep virtual
amplitudes; however, the parameterization thereof is effectively
not under control. This is illustrated in the right panel of
Fig.~\ref{Fig-GPDans}, where we show the GPD at its cross-over
trajectory within the spectator model (solid) of
Ref.~\cite{HwaMue07}, the leading SO(3) partial wave approximation
in the Mellin-Barnes representation (dash-dotted), and a VGG-like
ansatz (dashed).  For the latter two models we have used a
factorized $t$-dependence, and the form factor and  PDF were
evaluated in the spectator model. There is one rule which GPD
model  builders like to follow; namely, they mostly avoid
invisible degrees of freedom (those that die out in the
forward limit). In other words, one relies on the
implicit assumption that the holographic principle holds true.
Exceptions to this rule are the pion pole contribution
\cite{ManPilRad98,GoePolVan01} and the so-called  $D$-term
\cite{PolWei99}, where the latter is entirely related to the
subtraction constant in the dispersion relation \cite{Ter05}. In
the SO(3) $t$-channel partial wave expansion both of these
exceptions are assigned to $J=0$ contributions, as shown for the
subtraction constant in Sect.~\ref{Sec-Pha}. In the context of the
parton model \cite{Wei72}, and within its field theoretical model
realization for forward Compton scattering \cite{BroCloGun71},
such  a $J=0$  term relates low- and high-energy physics. We
emphasize that there are various possibilities to embody these $J=0$
terms into GPD models, see, e.g.,
Refs.~\cite{PolWei99,BelMueKirSch00,Ter01,TibDetMil04,KumMuePas07,HwaMue07},
and they might lead to rather different ad hoc GPD `predictions'.

\subsubsection{Ad hoc GPD phenomenology}

The qualitative failure of the VGG  model in describing  virtual
electroproduction of $\rho^0$ mesons in the resonance, i.e., large
$x_{\rm Bj}$, region \cite{GuiMor07} is partly related to its ad
hoc $t$-dependence. Certainly, here one has to be concerned about
the applicability of perturbative QCD (and, by the way, also about
the onset of the Regge regime) and so one might expect in this region
both large radiative and  power suppressed corrections. However, since the Feynman
mechanism is somehow incorporated in the GPD formalism, we might expect that the
$t$-dependence of the hadronic amplitude is qualitatively embodied
in the GPD; see Ref.~\cite{GolKro05,GolKro07} for a perhaps more
appropriate realization of the GPD formalism in this kinematics.
The $t$-dependence of the cross section diminishes in
the large $x_{\rm Bj}$ region \cite{GuiMor07}, which is consistent with
the aforementioned spectator GPD model \cite{HwaMue07}.%
\footnote{We thank the authors of Ref.~\cite{GuiMor07}, asking the
question ``GPDs or not GPDs?'',  for clarification. Their conclusions about GPD models
within DD representation refer only to  specific models,
considered there. In particular the statement {\em ``the very flat $t$ slopes
observed can only arise from GPD contributions not constrained by
the FF \emph{[form factor]} sum rule''}  is not meant to be valid in general. Indeed,
simple spectator GPD models, formulated in DD representation and
satisfying the form factor sum rule,  explain the observation. }
The notorious normalization problem  of the
cross section in this kinematical region is to LO approximation
entirely  related to the end-point behavior of the VGG model. It
was cured in Ref.~\cite{GuiMor07} by a $D$-term inspired addendum,
concentrated in the central GPD region and vanishing in the
forward kinematics. We will also
illustrate in Sect.~\ref{SubSec-RevGPD} that one can easily
construct GPD ans\"atze which could fix the normalization problem
in a different manner. In fact, it is demonstrated in Fig.~\ref{Fig-GPDans}(b),
that a  simple spectator model (solid curve) possesses in
comparison to a VGG-like ansatz (dashed curve) an enhancement
effect in the large momentum fraction region.

The DVCS amplitude is accessible in the deep virtual
electroproduction of photons, where due to the interference of the
bremsstrahlung and DVCS processes one can also reveal the phase of
the DVCS amplitude. The VGG model%
\footnote{
We would like to
emphasize that the `easy to tune' options in GPD codes indirectly
suggest the possibility of easy phenomenological GPD interpretation;
however, we fear that this can be misleading.
For instance, it was found in Ref.~\cite{Airetal08} that the
$D$-term within the VGG model is disfavored from the HERMES DVCS
measurements. Since the $\chi$QSM \cite{GoePolVan01,Goeetal07} and
now also lattice calculations, e.g., in Ref.~\cite{Hagetal07},
indicate that the $D$-term is sizeable, our
interpretation of this  statement is that the VGG model, the
incorporation of the $D$-term equivalent part (see Sect.~3.3) in the model
or a combination of both, is
improper. Also, the model dependent constraints on the quark
angular momentum in Ref.~\cite{Airetal08} illustrate that the
model dependence is not under control.
}
also fails to describe the beam
spin asymmetry and polarized cross section for unpolarized proton
in the fixed target JLab kinematics, see e.g.,
Ref.~\cite{Cametal06}, since here it does not offer the
possibility to decrease the normalization of the resulting
amplitudes, compared to the basic GPD ansatz.

The normalization problem
was already studied in Refs.~\cite{BelMueKir01,KirMue03} and to
some extent cured by reducing the amount of sea quark
contributions (introducing so-called model A or B, which, however,
have an unrealistic $t$-dependence). In fact, adjusting
the normalization of the DVCS amplitude by suppression of sea
quark contributions%
\footnote{\label{foo-AhmHonLiuTan}
Sea quark contributions and in
our understanding also valence singlet quark ones are neglected in
the DVCS results of Ref.~\cite{AhmHonLiuTan07}.
} makes it
possible to describe existing DVCS data, related to beam spin
asymmetries or polarized cross section measurements. Even
semi-quantitative discussions \cite{BelMueKir01,KirMue03} seem to
provide empirical understanding, when employed within the impulse
approximation in a prediction of beam spin asymmetry for DVCS on
nuclei \cite{Ell07}. If this `prediction' turns out to be in
disagreement with increasingly precise experimental data, it will
be exciting to employ the GPD formalism as a tool to explore the
interplay of nucleon and partonic degrees of freedom beyond the
impulse approximation.

The minimal(ist) version of the `dual'
parameterization \cite{PolShu02} has been confronted with DVCS
data in Refs.~\cite{GuzTec06} and \cite{PolVan08} and describes
asymmetries and polarized cross sections. Finally, all
these GPD models fail to describe the unpolarized DVCS cross
section, measured by the Hall A collaboration at JLab
\cite{Cametal06}.

The understanding of GPDs in collider kinematics of the H1
and ZEUS collaborations
\cite{Adletal01,Chekanov:2003ya,Aktas:2005ty,Aaretal07}
was in poor state, too. Claims that the  DVCS
cross section can be described within published GPD models to
LO accuracy are based on either a problematic implementation of
the $t$-dependence \cite{BelMueKir01,GuzTec06},%
\footnote{
Since the $t$-dependence of the sea quark GPD is unrealistic,
we rejected the model of Ref.~\cite{BelMueKir01}  for
some time. However, it describes various DVCS data and  we
employ it for error estimates in Sect.~4.2.
The description of data in
Ref.~\cite{GuzTec06} arises from a GPD model that induces an
unrealistic shrinkage effect of the diffractive forward peak.
} 
violation of the
perturbatively predicted scale change \cite{GuzTec06,FreMcDStr02},%
\footnote{
The incorporation of the ${\cal Q}^2$-dependent $t$-slope (32) in
Ref.~\cite{FreMcDStr02}, also adopted in Ref.~\cite{GuzTec06},
violates the LO scale change prediction.
}
or violation of basic GPD properties \cite{FreMcDStr02}.%
\footnote{ \label{foo-FFS}
The aligned jet GPD model \cite{FreMcDStr02} is based on
a truncated Taylor expansion in $\eta$ for all GPD moments.
The inconsistency of
the proposal \cite{FreMcDStr02}  was revealed in
Ref.~\cite{DieIva07}.
According to Eq.~(\ref{eq:Fnjint}), the moments in the `full'
model, evaluated from Eq.~(21, \cite{FreMcDStr02}), are
$\eta$-independent and the resulting GPD is  the
PDF decorated with a $t$-dependence.
}

We stress that the physical content of the LO normalization problem
was known to the authors of Ref.~\cite{FreMcDStr02}.
Although they started to analyze the model problems within the
RDDA, they finally gave up on the DD
representation. Armed with the aligned jet model, a reasonable
physical picture, however, violating Lorentz covariance, they
came up with an inconsistent GPD model (see footnote
\ref{foo-FFS}), which must be rejected. After restoring Lorentz
covariance in their model one simply finds a PDF which is
decorated with $t$-dependence --- a special case of the RDDA.

The  normalization problem can be naturally overcome by inclusion
of radiative corrections as advocated in
Refs.~\cite{FreMcD01a,FreMcD01c}. However, it  was at the same
time observed that GPD ans\"{a}tze within different PDF
parameterizations, describing the same set of unpolarized 
DIS data, result in rather different DVCS
predictions \cite{BelMueNieSch99}. This is quite
understandable if one remembers that already  the description of
the unpolarized DIS structure functions is a fine tuning problem.
There the perturbative expansion of the evolution is unstable in
the small $x_{\rm Bj}$ region, see discussion in
Ref.~\cite{VogNee00}, and the same thing now happens in the DVCS
case \cite{KumMuePas07}. We emphasize that this does not
disqualify the perturbative approach to DIS and DVCS at small
$x_{\rm Bj}$, because the evolution operator is universal, i.e.,
process independent. Also a resummation procedure might stabilize
the perturbative results; for a discussion see, e.g.,
Ref.~\cite{For05}. In particular, it will be phenomenologically
important for a perturbative GPD analysis of the deep virtual
vector meson electroproduction, see Ref.~\cite{Iva07} for details.

\subsubsection{A  fitting procedure}

The ad hoc GPD ansatz approach is surely not a proper framework
for revealing the partonic content of the proton. Insight
in the perturbative GPD approach is needed, too. We started to
explore an alternative framework  for DVCS in the small $x_{\rm
Bj}$ kinematics \cite{Mue06,KumMuePas07,KumMuePasSch08} up to
next-to-next-to-leading order
\cite{BelMue97a,BelMue98c,BelFreMue99,Mue05a,KumMuePasSch06} and
would like here to present what we have learned. Since radiative
corrections to the DVCS amplitude (or DIS structure function $F_T$)
can be essentially absorbed by a scheme change \cite{BelMueKir01},
the description of experimental DVCS data at LO accuracy is the
{\em indispensable key} to a GPD understanding.

\begin{figure}[t]
\begin{center}
\mbox{
\begin{picture}(250,150)(0,0)
\put(70,115){\small (a)}
\put(-110,0){\insertfig{7.5}{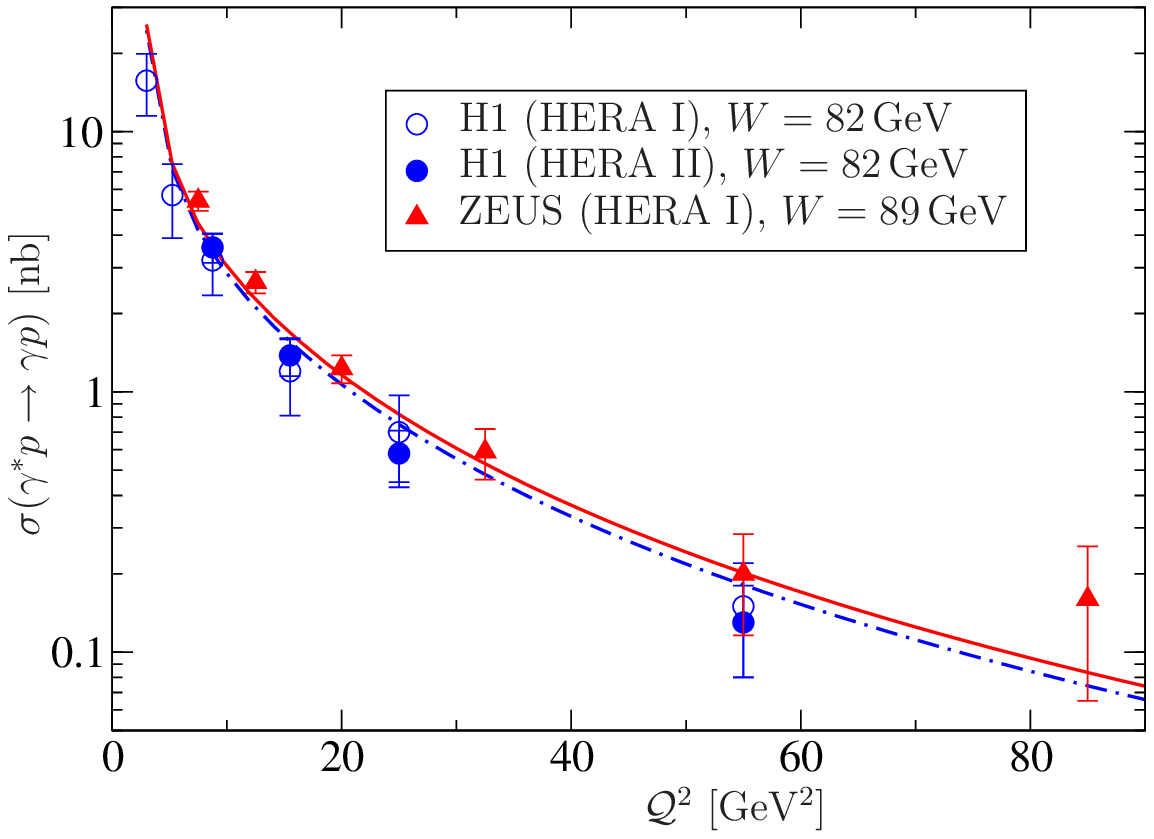}}
\put(330,115){\small (b)}
\put(150,0){\insertfig{7.5}{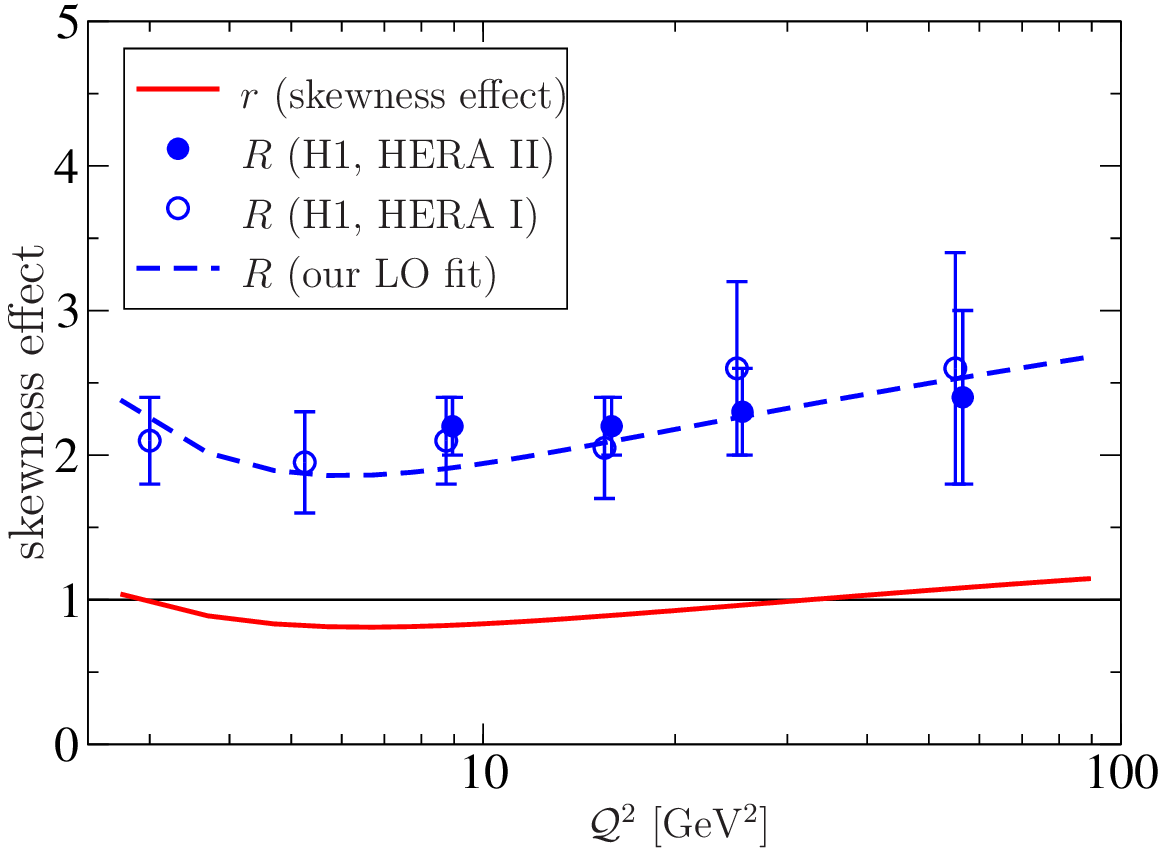}}
\end{picture}
}
\end{center}
\vspace{-2mm}
\caption{ \label{Fig-FIT}
The left panel shows a preliminary LO fit result
\cite{KumMuePasSch08}  to the HERA DVCS data set versus the photon
virtuality ${\cal Q}^2$: ZEUS \cite{Chekanov:2003ya} with $W=89\,
{\rm GeV}$ (triangle, solid curve)  and H1 with $W=82\, {\rm GeV}$
within the  HERA I \cite{Aktas:2005ty} (empty circle, dash-dotted
curve) and HERA II \cite{Aaretal07} (full circle, dash-dotted
curve) run. In the right panel we show the skewness ratio $R$,
defined  in Ref.~\cite{Aaretal07} as observable,  (dashed  curve) to
LO approximation, i.e., $R\approx H(x,x,t=0,{\cal
Q}^2)/H(2x,0,t=0,{\cal Q}^2)$ for fixed $W=82\, {\rm GeV}$, and
the ratio $r\equiv H(x,x,t=0,{\cal Q}^2)/H(x,0,t=0,{\cal Q}^2)$
for fixed $x=10^{-3}$ (solid curve). Note that $R \approx
2^{\alpha(0,{\cal Q}^2)}\, r$ with $\alpha(0,{\cal Q}^2\sim
4\,{\rm GeV}^2) \sim 1.2$.
}
\end{figure}
In the small $x_{\rm Bj}$ kinematics the functional form of the
amplitude simplifies drastically and the reggeized $t$-channel
view gives guidance for modelling. This is a pure empirical
recipe, where one has to replace the classical pomeron trajectory
by an effective one. The effective pomeron intercept is an
integral part of the universal GPD and it is considered to be
skewness independent, i.e., the GPDs or DVCS amplitude  have the
same intercept as the corresponding PDFs or DIS structure
function, respectively. This is supported by both a diagrammatical
$t$-channel ladder analysis \cite{FraFreGuzStr97} and
the experimental data.

For a fitting routine it is crucial to have full control
over the normalization of the DVCS amplitude. This can be achieved
within invisible terms that die out in forward kinematics.
The use of conformal symmetry enabled us to study the GPD approach
up to next-to-next-to-leading order in perturbation theory
\cite{Mue05a,KumMuePasSch06,KumMuePas07}. The numerical advantages
of the Mellin-Barnes representation then allow for a simultaneous
fitting procedure of DVCS and DIS data. The essential lesson from
our GPD fits is that a leading SO(3) partial wave approximation
(minimalist `dual' model) does not work in the small $x_{\rm Bj}$
regime at LO, but only at perturbative next-to-leading order or
beyond \cite{KumMuePas07}. Within the conformal partial wave
expansion, a more flexible parameterization of GPDs has been worked out
by several researchers; however, it is only partly published
\cite{tal07, Nor07, Pol07, KumMuePasSch08}. After such an
improvement, the experimental data for collider kinematics can be
fitted at LO accuracy either within the inclusion of the
next-leading SO(3) partial wave \cite{tal07} or a flexible model
dependent SO(3) partial wave resummation \cite{KumMuePasSch08}. A
preliminary fitting of results to the DVCS cross section, measured by
the H1 \cite{Aktas:2005ty,Aaretal07} and ZEUS
\cite{Chekanov:2003ya} collaborations, versus the photon
virtuality ${\cal Q}^2$ \cite{KumMuePasSch08} is displayed in the
left panel of Fig.~\ref{Fig-FIT}. Our result is that the dominant
sea quark GPD at low $x$ shows to LO accuracy almost no skewness
effect \cite{KumMuePasSch08}; see the solid curve in the right panel
of Fig.~\ref{Fig-FIT}. We should here give attention to a
different GPD interpretation of the same DVCS data. Namely, the
measured ratio $R$ \cite{Aaretal07} and our corresponding GPD fit
(dashed line), might be viewed as a large skewness effect.
Actually, using a better suited $r$ function, we reinterpret this as
(almost) no-skewness effect. Such a vanishing skewness effect
was for a lower resolution scale advocated in
Ref.~\cite{FraFreGuzStr97}, but it turns out that it holds  for
the experimental lever arm of $ \sim 3 \cdots 80\, {\rm GeV}^2$
\cite{Aaretal07}. In the published GPD models this no-skewness
property was either never present or, if it was taken at low
input scale \cite{FraFreGuzStr97}, the ansatz spoils it due to the
evolution \cite{FraFreGuzStr97}. Unfortunately, the large
evolution effect was propagated as a `prediction'
\cite{ShuBieMarRys99}, it is `naturally' implemented in
approximated versions of conformal partial wave GPD
representations, and it occurs  in a weaker form for RDDA, see
Ref.~\cite{DieKug07a}. We add that in the phenomenology of
diffractive vector meson production it is often taken for granted
that the  GPDs possess a positive non-zero skewness effect, e.g.,
in Ref.~\cite{MarRysTeu99}.

\subsubsection{Lessons from failure and success: a speculative GPD picture}
\label{SubSecLes}

{From} a phenomenological failure of present GPD models one can
mainly conclude that its ad hoc realization was improper. It is
senseless to criticize GPD representations in general, which will
just lead to a step backwards. GPD features which have been
understood in one representation can be incorporated in the other
representation, too.  Furthermore, from the confrontation of
experimental findings and present GPD models one cannot provide a
reliable partonic interpretation, in particular for the $D$-term
equivalent part and the quark angular momentum. The critique of the
perturbative approach offered so far from the outcome of such
a confrontation should be considered as speculation. In particular,
from what was said above for DVCS, it follows that a perturbative
GPD analysis for deeply virtual vector meson electroproduction at
small $x_{\rm Bj}$ also has to be reconsidered and extended to
next-to-leading order.

The failure of the mainstream ad hoc GPD models should be an
incentive to learn the lessons and find improved versions.  This
has been achieved to some extent with respect to the interplay of
momentum fraction and $t$-dependence
\cite{MukMusPauRad02,TibDetMil04,HwaMue07}. Pragmatical solutions
to fit data within the GPD formalism are obvious, namely, one has
to exploit degrees of freedom that are invisible in the zero
skewness case, e.g., more flexible profile functions.  This
sounds simple; however, it requires some technical effort and,
more importantly, GPDs are intricate functions and so one can
easily get `lost'. Even in the in some sense much simpler case of
small $x_{\rm Bj}$ kinematics, three different groups of authors
attempted in a period of five years or so to go for a GPD interpretation of
the ZEUS/H1 DVCS data, where two of them gave up basic (and
realistic) GPD properties. In our opinion the message from that
for the analysis of fixed target experiments is that one should
first utilize dispersion relation techniques, to pin down the GPD
on its cross-over trajectory.

Nevertheless, we attempt in the following to catch the possible
qualitative features of a `more realistic' GPD at its cross-over
trajectory from the phenomenological findings and we would like to
share our speculations on how it might look, away from the cross-over
trajectory. These unorthodox considerations are disputable and we
consider the outcome only as a working hypothesis.

Let us briefly summarize the phenomenological findings.  Present
ad hoc GPD models, perhaps apart from the VGG one, are in the
valence region consistent with (beam) spin asymmetry and polarized
cross section measurements. If they are tuned to experimental data
they might possess some kind of predictive power for certain
observables. There is now a clear understanding of the dominant
sea quark GPD  in the small $x_{\rm Bj}$ region, namely, the
skewness effect to LO accuracy is compatible with zero. The
extension of such an ansatz to larger values of $x$ induces a
wrong scale dependence. The large size of the unpolarized DVCS
cross section in the valence region is hardly understandable
within present GPD models.  The virtual $\rho^0$
electroproduction data  \cite{GuiMor07} in the large $x_{\rm Bj}$
region are consistent with respect to the $t$-dependence of simple
spectator models. A precise quantitative
understanding of the cross section is for various reasons rather
challenging.  We add that the employment of unitarity  constraints
in a Regge inspired model analysis \cite{Lag07} indicates that
for DVCS already the valence region cannot be correctly described
with a few reggeized $t$-channel exchanges.

Since the real part of the DVCS amplitude can be determined from a
dispersion relation, a GPD ansatz at the cross-over trajectory
must simultaneously give the correct description for the imaginary
part of the amplitude in the whole phase space. This is certainly
not the case for present GPD models; however, this GPD aspect
can be straightforwardly analyzed using the growing amount of DVCS
data.

On qualitative grounds we conclude from the experimental
findings that a GPD on its cross-over trajectory is at small $x$
governed by the reggeized  $t$-channel view, where the skewness
effect might be zero, while at large $x$ there is a large skewness
effect and the $t$-dependence diminishes. Such a behavior is
supported by field theory inspired spectator models
\cite{MukMusPauRad02,TibDetMil04,HwaMue07}; see, e.g.,
the discussion in Ref.~\cite{HwaMue07}, where, however, the Regge
behavior at small $x$ is absent. Implementing Regge-behavior from
the $s$-channel view, along the lines of Ref.~\cite{LanPol71},
yields at $t=0$ a Regge `improved' RDDA that possesses an
experimentally unfavored large skewness effect at small $x$. To
find a more realistic behavior it is perhaps more appropriate to
have the $t$-channel view by considering the diagrammatical
ladder, as started in Ref.~\cite{FraFreGuzStr97}. That this is a
reliable concept has been demonstrated for PDFs, see, e.g.,
Ref.~\cite{ErmGreTro00}.

We finally mention that due to the evolution the GPD value in the
outer region is reduced and `flows' into the central region to
`fill' it, i.e., yielding a growing value of the GPD. {From}
the experimental observation that the skewness effect at very
small $x$ `generically' does not depend on the resolution scale,
one might draw a speculative conclusion. Namely, loosely spoken,
the `flow' into the central region has already stopped and the GPD
in this region is mostly `filled' at a resolution scale of about
2 GeV or perhaps even less. This seems to fit the qualitative
experimental findings at large $x_{\rm Bj}$, indicating a `big'
GPD contribution in the central region \cite{GuiMor07}. Utilizing
the evolution equation, this pictorial speculation can be exactly
formulated as a mathematical problem; however, this is beyond the
scope of the paper.

\section{GPD duality and  families of GPD sum rules}
\label{Sec-DuaSumRul}

In this section we deliver the tools that allow for a
straightforward GPD analysis of  the photon electroproduction data in
fixed target kinematics. Thereby, we adopt well-known techniques,
utilized in the phenomenology of on-shell processes and combine
them with the duality property of GPDs
\cite{MueSch05,KumMuePas07}.

\subsection{Preliminaries}

To address the problems, we consider  virtual Compton scattering
off a proton in the generalized Bjorken region
\cite{MueRobGeyDitHor94}:
\begin{eqnarray}
\gamma^\ast(q_1)\; p(p_1) \to  \gamma^{(\ast)}(q_2)\; p(p_2)\,,\qquad
Q^2= -(q_1+q_2)^2/4 \to \infty\,,
\end{eqnarray}
where the momentum transfer squared $t\equiv \Delta^2 = (p_2-p_1)^2$ and
the two scaling variables
\begin{eqnarray}
\label{Def-ScaVar}
\xi = \frac{Q^2}{P\cdot q}\,, \quad \eta = - \frac{\Delta\cdot
q}{P\cdot q} \, ,
\end{eqnarray}
are fixed. Here $P=p_1+p_2$, $\Delta=
p_2-p_1$ and $q= (q_1+q_2)/2$. We shall always take the values of
$\xi$ and $\eta$ to be positive. In the twist-two approximation
the crossing-invariant (i.e., for $s\leftrightarrow t$) photon
asymmetry parameter is given by their ratio:
\begin{eqnarray}
\vartheta = \frac{q_1^2-q_2^2}{q_1^2+q_2^2}\simeq
\frac{\eta}{\xi}\,.
\end{eqnarray}
The  amplitude is parameterized in terms of the Compton form
factors (CFFs), which in perturbative LO, Born or hand-bag
approximation, read for leading twist-two:
\begin{eqnarray}
\label{Def-AmpLO} {\cal F}(\xi,\vartheta,t,Q^2) \stackrel{\rm
LO}{=} \int_{-1}^{1}\! dx\; \left(\frac{1}{\xi-x-i\epsilon} \mp
\frac{1}{\xi+x-i\epsilon}\right) F(x,\eta=\vartheta
\xi,t,\mu^2=Q^2)\,.
\end{eqnarray}
Here the scale $\mu^2$ is a factorization scale at which the
hard-scattering part and GPDs are factorized and it is equated to
the characteristic scale $Q^2$ of the process (the common choice). The
explicit dependence of the hard-scattering part on $\mu^2$ appears
at next-to-leading order. In our shorthand notation  CFFs will be denoted ${\cal
F}$ and the GPDs $F$, cf.~Ref.~\cite{BelMueKir01}:
\begin{eqnarray}
\label{Rel-Ima2GPD} {\cal F}=
\genfrac{\{}{\}}{0 pt}{}{{\cal H},{\cal  E}}
{\widetilde {\cal  H}, \widetilde {\cal  E} } (\xi,\vartheta,t,Q^2)\,, \quad
F= \genfrac{\{}{\}}{0 pt}{} {H,E}
{\widetilde H, \widetilde E }(x, \eta=\vartheta
\xi,t,\mu^2) \quad \mbox{for parity} \quad
\genfrac{\{}{\}}{0 pt}{}{\rm even}{\rm odd} \,.
\end{eqnarray}
The sign $\mp$ on the r.h.s.~of Eq.~(\ref{Def-AmpLO}) refers to
the symmetry with respect to $x$. This convention will be
consistently used below without an explicit reference to it. Note
that antisymmetry (symmetry) with respect to $x$ corresponds to
parity even (odd) CFFs, which also have signature $+1 (-1)$. This
definite signature originates from symmetry under the exchange of
$s$- and $u$-channel. Here we did not indicate fractional charge
squared factors, which can easily be restored by replacing $F$
with $\sum_{a=u,d,\cdots} Q_a^2\, {^a\!F}$. The GPDs $F(x,
\eta,t,Q^2)$ are real-valued functions that are symmetric in
$\eta$ \cite{Ji98}, and their $j$th Mellin moments are
polynomials in $\eta$ of order $j+1$ (polynomiality property). To
be on safer ground concerning the analyticity hypothesis, we
consider in the following only the Euclidean region, i.e.,
$|\vartheta| \le 1$. This is ensured by the condition $\xi>\eta$.
Unfortunately, apart from the contribution to virtual two-photon
exchange corrections \cite{AfaCar05}, the CFFs are only accessible
in DIS ($\vartheta=0$, i.e., $\eta=0$) and DVCS ($\vartheta=1$,
i.e., $\eta=\xi$). In the former case we extract PDFs, equal to
GPDs with $\eta=0$ and $t=0$, and in the latter case we find the GPDs
on the cross-over trajectory $x=\eta$. For pedagogical reasons,
however, we shall first do a gedanken experiment and suppose that
the Compton amplitude can be `measured' in the whole Euclidean
region.

The obvious part of the answer to the  question `What GPD
information can be revealed from experiment?' is that the
imaginary part of the CFF (\ref{Def-AmpLO}) is given by the GPD
combination:
\begin{eqnarray}
\label{InvRel} F^\mp (x,\eta,t,Q^2) \equiv F(x,\eta,t,Q^2)\mp
F(-x,\eta,t,Q^2)  \stackrel{\rm LO}{=} \frac{1}{\pi}\Im{\rm
m}{\cal F}(\xi=x,\vartheta= \eta/x,t,Q^2) \,.
\end{eqnarray}
This formula tells us that by varying the photon asymmetry parameter, i.e.,
$0 \le \vartheta\le 1$, the GPD is scanned in the outer region
$$  0 \le \eta \le x=\xi  \le 1.$$
This is analogous to DIS, where the structure functions are to LO
equal to PDFs. The r.h.s.~in Eq.~(\ref{InvRel}) and so the GPD
$F^\mp(x,\eta,t,Q^2)$ with $\eta \le x $ might be viewed as the
spectral function for this Compton scattering process, given by
the $s$-channel cut. Hence, the outer region of the GPD
$F(x,\eta,t,Q^2)$ with $\eta\le x$ is expressed by the difference
or sum of $s$- and $u$-channel cuts. In accordance with the
standard definition of the PDFs, the GPD (\ref{InvRel}) is understood
for $\eta \le x $ as the combination $F^\mp(x \ge
\eta,\eta,\cdots)=\left[ {^q\!F} +
{^{\bar{q}}\!F}\right](x,\eta,\cdots)$ of quarks and antiquarks,
where
\begin{eqnarray}
\genfrac{\{}{\}}{0pt}{}{  {^q\!F} }{   {^{\bar{q}}\!F}
}(x,\eta,\cdots) &\!\!\! =\!\!\!& \genfrac{\{}{\}}{0pt}{}{ F(x,\eta,\cdots)
}{ \mp
F(-x,\eta,\cdots)} \quad  \genfrac{\{}{.}{0pt}{}  {\mbox{quark} }{
\mbox{antiquark} }  \quad \mbox{for} \quad x \ge \eta\,.
\end{eqnarray}
The definite symmetry of $F^{\mp}$ GPDs under reflection $x\to -x$
allows us to extend them uniquely to negative values $x \le -\eta$.
The support extension into the central region $x \in [-\eta,\eta]$
should be done in such a way that the polynomiality condition is
satisfied. As we shall see in Sect.~\ref{SubSec-DecDua} below,
analyticity guarantees that this procedure is unique.

\subsection{Sum rules}

To get deeper into the problem, we consider the CFFs as
holomorphic functions of Mandelstam variables which satisfy the
Schwartz reflection principle. Furthermore, we assume that for
fixed (negative) value of $t$ in the Euclidean region the CFFs
possess only elastic poles, as well as $s$- and $u$-channel
discontinuities on the real axis. Hence, CFFs satisfy a single
variable dispersion relation, where $t$ and the photon
virtualities are considered as fixed variables; see, e.g.,
Refs.~\cite{FraFreGuzStr97,Che97,Ter05,KumMuePas07,DieIva07}.
Taking into account that they have a definite signature, this can be
equivalently rewritten in terms of the crossing-symmetric variable
$\nu = (s-u)/4M$, related to the energy of the initial photon, or
in terms of the scaling variable $\xi= Q^2/2 M\nu$, originally
defined in Eq.~(\ref{Def-ScaVar}):
\begin{eqnarray}
{\cal F}(\xi,\vartheta,t,Q^2) = \frac{1}{\pi}\int_{0}^{1}\!
d\xi^\prime \left(\frac{1}{\xi-\xi^\prime - i \epsilon} \mp
\frac{1}{\xi+\xi^\prime- i \epsilon} \right) \Im{\rm m}{\cal
F}(\xi^\prime-i 0,\vartheta,t,Q^2) +{\cal C}_{\cal F}(\vartheta,t,Q^2) \,.
\label{Def-DisRel}
\end{eqnarray}
For convenience, we here set the upper integration limit $\xi_{\rm
cut}=\xi_{\rm pol} = 1/(1+t/4 Q^2)$ to 1, according to the
Bjorken limit, so that the elastic pole is included in the
imaginary part, rather than in the subtraction constant. We
emphasize that the correct upper bound can easily be restored by
the substitution $1\to 1/(1+t/4 Q^2)$. The subtraction constant is
fixed by the value of the CFFs at $\xi=\infty$.
It appears in the CFFs $\cal H$ and $\cal E$, vanishes
for the CFFs $\widetilde{\cal H}$ and $\widetilde{\cal E}$, and is
perturbatively predicted to be zero for the combination ${\cal H}+
{\cal E}$, see Ref.~\cite{KumMuePas07}. Thus, we can define:
\begin{equation}
{\cal C}(\vartheta,t,Q^2) \equiv {\cal C}_{\cal E}(\vartheta,t,Q^2)= {\cal E}(\xi=\infty,\vartheta,t,Q^2) \,, \qquad
{\cal C}_{\cal H} (\vartheta,t,Q^2)  =   - {\cal C}(\vartheta,t,Q^2) \,,
\label{Def-SubCon}
\end{equation}
where ${\cal C}_{\widetilde{\cal H}} ={\cal C}_{\widetilde{\cal E}} = 0$.

The dispersion relation (\ref{Def-DisRel}) states that both the
imaginary and real part of CFFs contain, up to a possible
subtraction constant, the same information:
\begin{eqnarray}
\Re{\rm e}{\cal F}(\xi,\vartheta,t,Q^2) =
\frac{1}{\pi} {\rm PV}\int_{0}^{1}\!
d\xi^\prime \left(\frac{1}{\xi-\xi^\prime} \mp
\frac{1}{\xi+\xi^\prime} \right) \Im{\rm m}{\cal
F}(\xi^\prime-i 0,\vartheta,t,Q^2) +{\cal C}_{\cal F}(\vartheta,t,Q^2)
\, .
\label{Def-DisRel1}
\end{eqnarray}
Let us loosely have a look at this statement without discussing any
mathematical refinements. After extending the support of $\Im{\rm
m}{\cal F}(\xi,\ldots)$  to $|\xi| >1$ and using the symmetry
property, we express Eq.~(\ref{Def-DisRel1}) as a (singular)
Fredholm type integral equation of the first kind with constant limits
($\pm \infty$). Briefly speaking,  the real part of ${\cal
F}(\xi,\ldots) -{\cal C}(\ldots)$ is the Hilbert transform of its
imaginary part. The inversion reads then
\begin{eqnarray}
\Im{\rm m}{\cal F}(\xi,\vartheta,t,Q^2) =
\frac{1}{\pi} {\rm PV} \int_{-\infty}^\infty\! d\xi^\prime\, \frac{1}{\xi^\prime-\xi}
\left[
\Re{\rm e}{\cal F}(\xi^\prime,\vartheta,t,Q^2)-{\cal C}_{\cal F}(\vartheta,t,Q^2)
\right]
\,.
\label{Def-InvDisRel}
\end{eqnarray}
Alternatively, by again making use of the spectral condition $\Im{\rm
m}{\cal F}(\xi,\ldots) =0$ for $|\xi| >1$, we can take the
solution of the  Fredholm type integral equation with constant limits of
integration ($\pm 1$); e.g., from Ref.~\cite{ManPol98}:
\begin{eqnarray}
\Im{\rm m}{\cal F}(\xi,\vartheta,t,Q^2) =
\frac{\sqrt{1-\xi^2}}{\pi} {\rm PV} \int_{-1}^1\! d\xi^\prime\,
\frac{1}{\sqrt{1-{\xi^\prime}^2}}
\frac{\Re{\rm e}{\cal F}(\xi^\prime,\vartheta,t,Q^2)
}{\xi^\prime-\xi}\quad\mbox{for}\quad |\xi| \le  1
\,.
\label{Def-InvDisRel-1}
\end{eqnarray}
In conclusion, assuming the specific analytic properties, we can
state that the imaginary and real part of CFFs, having definite
signature and a vanishing  subtraction constant, have a one-to-one
correspondence in the `accessible` region $0\le \xi \le 1$. This
correspondence seems to become imperfect in the case of
non-vanishing subtraction constant (\ref{Def-SubCon}). As in
Ref.~\cite{KumMuePas07}, we shall argue below that this might not
be the case, if  the $\xi\to 0$ asymptote is given by Regge
behavior, as is commonly assumed.

In a real experiment one accesses only a limited phase space
region. To pin down CFFs in inaccessible regions one can again
employ their holomorphic property, but now expressed as a family of
sum rules. To give an example, we recall that Regge asymptotics of
on-shell amplitudes yield finite energy sum rules (FESRs), which
are widely used in phenomenology. Pragmatically, we suppose that
off-shell CFFs possess the high-energy asymptotic power-like
behavior $\xi^{-\alpha(t)}$, where $\alpha(t)$ is an effective
Regge trajectory, which might depend on $Q^2$. We introduce now a
duality parameter $\upsilon$ that divides the low- ($\upsilon <
\xi^\prime$) and high- ($\xi^\prime < \upsilon$) energy region in
the dispersion integral (\ref{Def-DisRel1}). Expanding in the
vicinity of $\xi=0$ yields FESRs%
\footnote{Assuming that $\Im{\rm m}{\cal F} \approx \pi
\sum_{\alpha} \beta_\alpha \xi^{-\alpha}/\Gamma(\alpha+1)$ is
approximately valid for $|\xi| \le \upsilon$ one straightforwardly
finds  for $\alpha >0$ and $n>0$: $S_n \approx  \sum_{\alpha}
\frac{\beta_\alpha}{\Gamma(\alpha+1)} \frac{ \upsilon^{-\alpha-n
}}{\alpha+n}$. To obtain the textbook form of $S_n$  for on-shell
amplitudes \cite{Col77}, one might now  rewrite the FESR
(\ref{Def-FESR}) in terms of the variables $\nu=Q^2/2 M \xi$ and
$N=Q^2/2 M \upsilon$  and take instead of the Bjorken limit the
on-shell one. To find the more general form, including $\alpha <0$,
one should follow the lines of Ref.~\cite{DolHorSch68}, where for $n=0$ the
$\alpha=0$ case has to be treated separately, e.g., as in
Ref.~\cite{CreDrePas69}.
}
\begin{eqnarray}
\label{Def-FESR} &&\!\!\!\!\int_\upsilon^1\! d\xi\, \xi^{-m-2}\,
\Im{\rm m}{\cal F}(\xi,\cdots) = \pi S_{m+1}(\upsilon,\cdots)\,,
\quad {\rm with}
\\
&&\!\!\!\! S_n(\upsilon,\cdots)= \frac{-2^{-1}}{n!}
\frac{d^{n}}{d\xi^{n}}\!\left[\Re{\rm e}{\cal F}(\xi,\cdots) -
{\cal C}_{\cal F}(\cdots) -\frac{1}{\pi} {\rm PV}\!\!
\int_{0}^{\upsilon}\!\! d\xi^\prime \left(\frac{1}{\xi-\xi^\prime
} \mp \frac{1}{\xi+\xi^\prime}\! \right) \Im{\rm m}{\cal
F}(\xi^\prime,\cdots)\! \right]_{\xi=0}. \nonumber
\end{eqnarray}
For odd (even) $m$ they are valid for signature $+1 (-1)$. We
emphasize that the $m=-1$ case deserves special attention. The
resulting FESR is closely related to what is  called  the inverse
moment. For real Compton scattering the form of this specific FESR
can be found in Ref.~\cite{CreDrePas69}.

The concealed part of the answer to the question `What GPD
information can be revealed from experiment?', arises  from the
analyticity of the amplitude, which ties the real and imaginary
parts by the single variable dispersion relation
(\ref{Def-DisRel}), advocated in Ref.~\cite{Ter05}. Plugging the
LO result (\ref{InvRel})  for the imaginary part into the
dispersion relation (\ref{Def-DisRel}), it becomes obvious that
the real part of the amplitude (\ref{Def-AmpLO}), up to a possible
subtraction constant, is also expressed by the outer GPD region:
\begin{eqnarray}
\label{AmpinGPD} \Re{\rm e}{\cal F}(\xi,\vartheta,t,Q^2)
\stackrel{\rm LO}{=} {\rm PV}\int_{0}^{1}\! dx
\left(\frac{1}{\xi-x} \mp \frac{1}{\xi+x} \right)
F^{\mp}(x,\eta=\vartheta x,t,Q^2) + C_F(\vartheta,t,Q^2)\,.
\end{eqnarray}
This `dispersion relation' tells us that for given $|\vartheta|
\le 1$  the variation of the real part with respect to $\xi$ is
governed by the value of the GPD on the trajectory $\eta=\vartheta
x$. Note that we have introduced a new symbol for the subtraction
constant fixed to LO:
\begin{equation}
 C_{F}(\vartheta,t,Q^2) \stackrel{\rm LO}{\equiv}
{\cal C}_{\cal F}(\vartheta,t,Q^2)
\, .
\end{equation}

\begin{figure}[t]
\begin{center}
\includegraphics[scale=0.9]{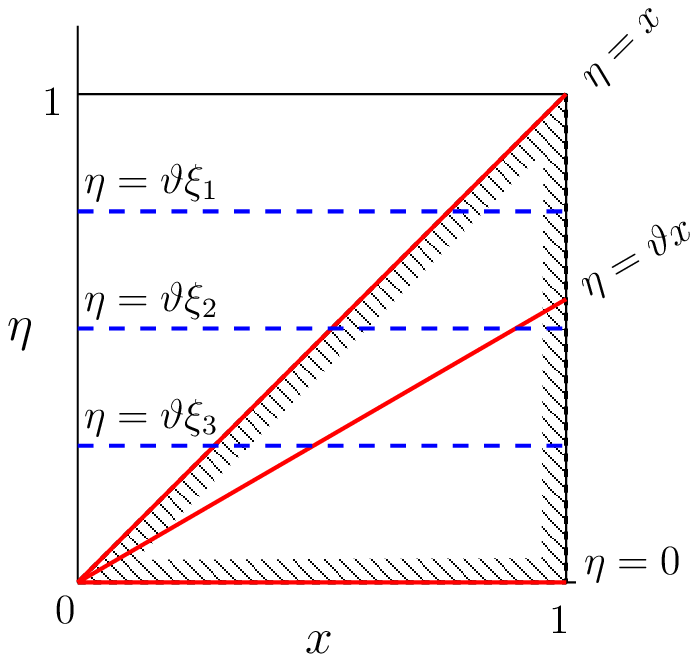}%
\hfill\includegraphics[scale=0.55,angle=-90,origin=br,clip]{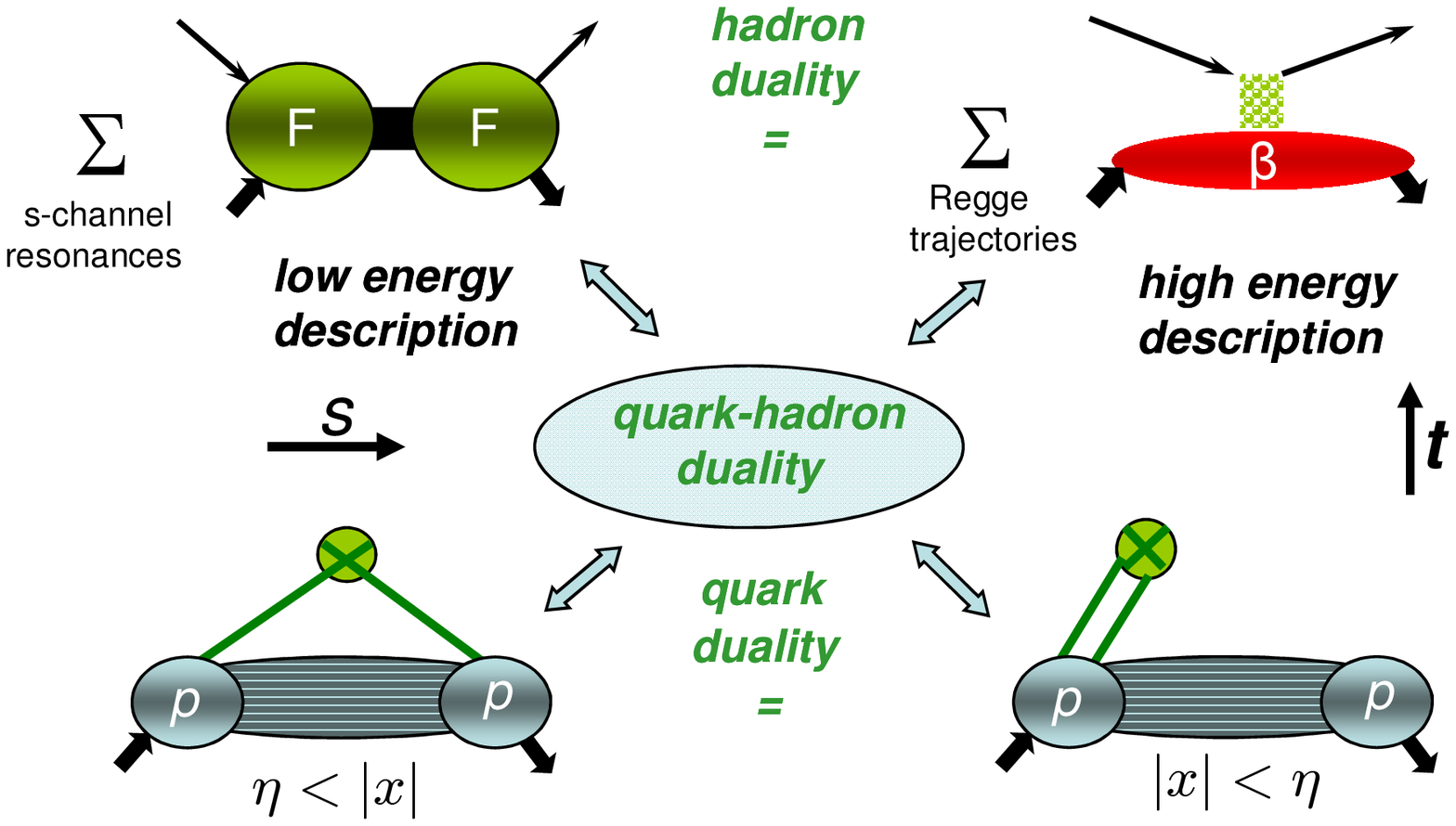}
\centerline{\hspace*{16ex}(a)\hfill(b)\hspace*{24ex}}
\end{center}
\caption{\label{Fig-Dua}
(a) The integration paths in the sum rule
(\ref{Def-GDPSR}) along the $\eta=\vartheta x$ (solid, red) and
horizontal $\eta=\vartheta\xi$ (dashed, blue) trajectories are
displayed. (b) Visualization of the approximate duality
relation of a GPD with the $s$- and $t$-channel view on the
hadronic amplitude. For on-shell amplitudes duality between the
$s$- and $t$-channel view has been conjectured at the end of the
sixties. Quark-hadron duality was conjectured in particular in DIS,
while modelling GPDs in the central region by the crossed
$t$-channel contribution was proposed in \cite{PolShu02}. }
\end{figure}

Compared to the LO formula (\ref{Def-AmpLO}), in which the whole
GPD support contributes to the real part, the observation,
manifest in Eq.~(\ref{AmpinGPD}), looks perhaps at the first
sight rather surprising. It is a consequence of analyticity and
the decomposition of the CFFs in terms of GPDs, which posses
definite  properties under Lorentz transformations, clearly
visible in the operator product expansion approach (OPE)
\cite{Che97,KumMuePas07}. The consequence is that the GPD in both
the central and the outer region know about each other and talk to
each other via the trajectory $\eta=\vartheta x$
\cite{GeyDitHorMueRob88,MueRobGeyDitHor94,MueSch05}. To obtain the
equality that governs this cross-talk, we combine formula
(\ref{AmpinGPD}) with the LO amplitude (\ref{Def-AmpLO}) and find
a \emph{family of GPD sum rules} (GPDSRs) \cite{Ter05,DieIva07}:
\begin{eqnarray}
\label{Def-GDPSR}
\int_0^{1}\! dx\;
\left(\frac{1}{\xi-x}\mp\frac{1}{\xi+x}\right)
\left[F^{\mp}(x,\eta=\vartheta \xi,t,Q^2)-F^{\mp}(x,\eta =
\vartheta x,t,Q^2)\right]=C_F(\vartheta,t,Q^2)\,,
\end{eqnarray}
parameterized by $\vartheta$. Here we have dropped the principal
value prescription, since the integrand can be considered as
integrable at $x=\xi$. The GPDSR family (\ref{Def-GDPSR}) is
visualized in Fig.~\ref{Fig-Dua}a. It relates
the integral along the $\eta=\vartheta x$ trajectory (solid) in the
outer GPD region to a family of integrals along the horizontal
$\eta=\vartheta\xi$ trajectories (dashed), traversing both inner
and outer regions. The variations of a spectator model GPD  on a
horizontal (dashed) and the cross-over trajectory (solid) are
displayed in the left panel of Fig.~\ref{Fig-GPDans}.

\subsection{Deconvolution and duality}
\label{SubSec-DecDua}

Let us suppose that the GPD in the outer region and the
subtraction constant are known from a `measurement'. Finding the
GPD in the central region can then be considered as a
deconvolution problem. Decomposing the integral regions into the
central and the outer one, one can view the GPDSR family
(\ref{Def-GDPSR}) as a Fredholm type integral equation of the
first kind, which allows one to solve the deconvolution problem. This
mapping is an integral transformation that requires the
continuation in the region $\eta>1$. Such an extension is already
defined in the DD representation%
\footnote{ The factor $(1-x)^p$ in our representation
(\ref{Def-DD})  is motivated by the form of the positivity constraints
\cite{Pob02} and appears in the model of Ref.~\cite{HwaMue07}. For
the GPD $E$ we have $p=1$ and thus a non-vanishing subtraction
constant. For GPD $H+E$ and the two parity odd GPDs $p=0$ and the
subtraction constant is absent.
}
\begin{eqnarray}
\label{Def-DD}
F(x ,\eta,t) = (1-x)^p
\int_0^1\!dy\int_{-1+y}^{1-y}dz\,  \delta(x-y-\eta\, z)f(y,z,t)\,.
\end{eqnarray}
One can verify that the (anti-) symmetrized GPDs within this
representation satisfy the GPDSR family (\ref{Def-GDPSR}). The
DD representation has been utilized in
Ref.~\cite{HwaMue07} to solve the deconvolution problem for a
simple spectator model, formulated in the light-cone wave function
overlap representation, see also
Refs.~\cite{MukMusPauRad02,TibDetMil04}.

Within the GPDSR family (\ref{Def-GDPSR}) the deconvolution
problem can also be solved by a constructive method that
explicitly avoids the continuation into the region $\eta>1$. This
method resembles the derivation of the standard factorization
formula (\ref{Def-AmpLO}) from the short distance OPE and the
dispersion relation (\ref{Def-DisRel}) in the Euclidean region,
see Refs.~\cite{Che97,KumMuePas07}.

We go now in reverse direction along the lines of
Ref.~\cite{KumMuePas07}. Expanding the integral
kernel in the GPDSR family (\ref{Def-GDPSR}) in
powers of $1/\xi$ relates integral Mellin moments,
$$
\int_{-1}^1 dx \, x^j F(x,\eta= \vartheta x,t,\mu^2)
\quad \mbox{for} \quad j \ge 0\,,
$$
of the GPD at the trajectory $\eta = \vartheta x$ to the usual Mellin
moments of the GPD with argument $\eta=\vartheta \xi $:
\begin{equation}
f_j(\eta,t,\mu^2)=\int_{-1}^1 dx \, x^j F(x,\eta,t,\mu^2)
\quad \mbox{for} \quad j \ge 0\,.
\label{eq:MelMom}
\end{equation}
The latter are defined in terms of the expectation values of local
operators with spin $j+1$. Lorentz covariance (or polynomiality
condition) enforces
\begin{equation}
\label{MelMom-pol}
f_j(\eta,t,\mu^2) = \sum_{\substack{n=0 \\ {\rm
even}} }^{j+1} f^{(n)}_j(t,\mu^2)\, \eta^n\,,
\end{equation}
where $f^{(j+1)}_j=0$ for $f=\{h+e,\widetilde h, \widetilde e\}$
(for details see, e.g., \cite{Die03a,BelRad05}). After reordering,
we equate the coefficients in front of $\xi^{-j-1}$
\begin{eqnarray}
\label{Def-Melmomxx}
\int_{0}^1\!d x\, x^j F^\mp(x,\vartheta x,t, Q^2)  =
\sum_{\substack{n=0 \\ {\rm even}} }^\infty \vartheta^n f^{(n)}_{j+n}(t,Q^2)
\,,
\end{eqnarray}
and thus
find the coefficients of the polynomial (\ref{MelMom-pol}):
\begin{equation}
f^{(n)}_j(t,Q^2) =
 \frac{1}{n!}   \frac{d^n}{d\vartheta^n} \int_{0}^1\! dx\,
x^{j-n} F^\mp(x, \vartheta x,t,Q^2) \Big|_{\vartheta=0}\quad \mbox{for}\quad j-n \ge 0\,,
\label{eq:Fnjint}
\end{equation}
where $j=1,3,5\cdots$ and $j=0,2,4\cdots$ for $-$ and $+$ type of
GPDs, respectively, and  $j-n \ge 0$ with $n$ even.
If a subtraction constant is present in ${\cal E}$, the
Mellin moments with $n=j+1$ appear and are evaluated from
\begin{equation}
e^{(j+1)}_j(t,Q^2) = 
\frac{1}{2} \frac{1}{(j+1)!}   \frac{d^{j+1}}{d\vartheta^{j+1}}
C(\vartheta,t,Q^2) \Big|_{\vartheta=0}\,.
\label{eq:Fnjint-C}
\end{equation}
Note that $h^{(j+1)}_j(t,Q^2)=-e^{(j+1)}_j(t,Q^2)$,
and that here, as in the rest of the paper, we
follow the convention (\ref{Def-SubCon}), i.e., $C \equiv C_E$ while $C_H=-C$.

All possible techniques are now at hand to construct the GPD from
the moments (\ref{MelMom-pol}), (\ref{eq:Fnjint}) and (\ref{eq:Fnjint-C}): standard inverse
Mellin transform, expansion with respect to conformal
eigenfunctions \cite{BelGeyMueSch97,Shu99,Nor00,MueSch05}, and
introduction of DD as generating function for the
Mellin moments \cite{MueRobGeyDitHor94,Rad98a}, see, e.g.,
Ref.~\cite{BelKirMueSch01}. All of them are based or related to
the uniqueness of an analytic continuation of the even or odd Mellin
moments, ensured by the Carlson theorem \cite{Car14}. Thereby, the
Mellin moments $e_j^{(j+1)}$ might either be considered connected
with a `regular part' of the GPD, e.g., as in Eq.~(\ref{Def-DD}),
or, alternatively, they can be included in the  $D$-term addenda
of a GPD \cite{PolWei99}. It is mostly a matter of taste and
convenience (as long as the whole GPD is considered as known),
which representation one prefers \cite{Ter01}. Only in the case
that the spectral function, i.e., $\Im{\rm m}{\cal F}$, possesses
unexpected analytic properties such as `invisible' terms, like $x
\delta(x)$, one has to treat the $e_j^{(j+1)}$ terms separately
from the DD part. In any case, after convolution
of the GPD with the hard-scattering part one has the
representation for the subtraction constant,
cf.~Eq.~(\ref{eq:Fnjint-C}):
\begin{eqnarray}
 C(\vartheta,t,Q^2)= 2\sum_{\substack{j=1 \\ {\rm odd}}}^\infty
\vartheta^{j+1} e_j^{(j+1)}(t,Q^2)\quad \mbox{with}  \quad  C(\vartheta=0,t,Q^2)
= 0\,,
 \label{eq:CsumFnn1}
\end{eqnarray}
where the normalization at $\vartheta=0$ is the perturbative
prediction.  This vanishing of the subtraction constant can be
considered as a consequence of the absence of a local and gauge
invariant spin-zero operator in the short distance OPE.

That the GPD in the outer region can be uniquely restored from its
knowledge in the central one is known and was again clearly
spelled out in Ref.~\cite{MueSch05}. This mapping  is (implicitly)
done in the conformal partial wave expansion approach by taking
appropriate boundary conditions for the partial waves; see
\cite{MueSch05,KirManSch05a} for a discussion. Hence, both regions
contain the same information and so they are dual to each other.

In anticipation of phenomenological applications we shall now
derive GPDSRs that have the form of FESRs. Note that GPDs and
their sum rules depend on the perturbative approximation and
scheme conventions that are used in the factorization approach.
Analogous to the the derivation of FESRs, we introduce the duality
parameter $\upsilon$, which divides the integral in the GPDSR
family (\ref{Def-GDPSR}) into the `low-' and `high-energy'
contribution. By expanding the integral kernel in
Eq.~(\ref{Def-GDPSR}) around $\xi=0$ we find the desired
GPDSRs:
\begin{eqnarray}
\label{Def-FEGDPSR}
\int_{\upsilon}^{1}\! dx\;
x^{-m-2}
\left[F^{\mp}(x,\vartheta x ,t,Q^2)- \sum_{\substack{n=0 \\ {\rm even}}}^{m+1} \frac{(\vartheta x)^{n}}{n!}
\frac{d^{n}}{d\eta^{n}}F^{\mp}(x,\eta,t,Q^2)\Big|_{\eta=0}\right] =
\delta S^{\mp}_{m+1}(\upsilon,\vartheta,t,Q^2)\,,
\end{eqnarray}
where  $m \ge -1$ is odd (even) for $-$ ($+$) type GPDs. The `high-energy' part $\delta S^{\mp}_{n}(\upsilon,\vartheta,t,Q^2)$
is obtained from the Taylor coefficients:
\begin{eqnarray}
\label{Def-FEGDPSR-S}
\delta S^{\mp}_{n} = \frac{-2^{-1}}{n!} \frac{d^n}{d\xi^n}
\! \left[
\int_0^{\upsilon}\! dx
\left(\!\frac{1}{\xi-x}\mp\frac{1}{\xi+x}\!\right)\!
\left[F^{\mp}(x,\vartheta \xi,\cdots)-F^{\mp}(x,
\vartheta x ,\cdots)\right]  -C_F(\cdots)\right]_{\xi=0}\!.
\end{eqnarray}
The l.h.s.~of the sum rule (\ref{Def-FEGDPSR}) relates for $x \ge \upsilon$ the GPD at the
trajectory $\eta =\vartheta\, x $ with its behavior in the
vicinity of $\eta=0$, where the strength of the skewness effect is
given by its r.h.s., i.e., it is
embodied in the `high-energy' constants $\delta
S^{\mp}_{n}(\upsilon,\vartheta)$.

We add that the `low-energy' and `high-energy' content might be
related by a more elegant, however, quite formal way, using
analytic (or canonical) regularization \cite{GelShi64}. This is
useful if one deals with GPD models formulated in Mellin space
with meromorphic functions. The GPD inherits the Regge-like
behavior of the CFFs, which we assume, and should not possess any
`invisible' terms, e.g., such as $x \delta(x)$. Under this assumption one can introduce
analytic regularization at $x=0$, indicated by $(0)$ as the lower
integration limit, in the sum rule (\ref{Def-GDPSR}) and expand
the integral kernel in the vicinity of $\xi=0$.  An example is
the evaluation of the subtraction constant from the `high-energy'
limit of Eq.~(\ref{Def-GDPSR}):
\begin{eqnarray}
\label{Cal-C-low}
C(\vartheta,t,Q^2)  = \lim_{\xi\to 0}
\int_{0}^1\! dx\, \frac{2 x}{x^2-\xi^2}
\left[E^-(x, \vartheta x,t,Q^2)-E^-(x, \vartheta \xi,t,Q^2)\right] \,.
\end{eqnarray}
It is obvious that interchanging the limit with the integration
could render a non-integrable singularity at $x=0$. Before
this interchange, one first has to introduce a regularization,
which does not spoil the normalization condition
$C(\vartheta=0,t,Q^2)=0$. In the absence of `invisible' terms, such as
$x \delta(x)$, we can utilize analytic  regularization and write, as in
Ref.~\cite{KumMuePas07}:
\begin{eqnarray}
\label{Cal-SubC-3}
C(\vartheta, t,Q^2) = \int_{(0)}^{1}\! dx\;
\frac{2}{x} \left[E^-(x,\vartheta x,t,Q^2)-E^-(x,0,t,Q^2)\right] .
\end{eqnarray}
It can easily be verified that this prescription, as well as
Eq.~(\ref{Cal-C-low}), is consistent with the  GPDSR
(\ref{Def-FEGDPSR}) for $m=-1$. Under the assumption of Regge
behavior, an analogous relation should hold for the imaginary part of
CFFs, too, as argued in Ref.~\cite{KumMuePas07} based on
knowledge of radiative corrections.

\section{Phenomenological applications}
\label{Sec-Pha}

\subsection{Strategy to access GPDs}
\label{Sec-Pha-Str}

As we saw, the imaginary part of the CFFs in the region $x=\xi \in
[0,1]$ is a spectral function, considered as `measurable' in our
Compton gedanken experiment for $|\vartheta| \le 1$. The real
part, arising from the dispersion relation (\ref{Def-DisRel}), is
then immediately known, too, and it must coincide with the `measured'
one in the region $x=\xi \in [0,1]$. By means of
Eq.~(\ref{Def-InvDisRel-1}), the logic can also be reversed:
having restricted `experimental access', i.e., knowing the
imaginary and real part of the CFFs only in a certain limited
interval, one can use the dispersion integral (\ref{Def-DisRel})
(and also Eq.~(\ref{Def-InvDisRel})) as a `local' sum rule, which
restricts the shape of the spectral function in the
`non-accessible' region. Such a restriction is possible even if one
knows the real part only at one given value of $\xi$. Various
forms of sum rules can be derived and we only concentrate on the
FESR ones, which relate the `low-energy' and `high-energy'
regions. `Measuring' the spectral function in dependence of
$\vartheta$, for $|\vartheta| \le 1$, allows us to solve the
deconvolution problem and to reveal the GPD. This can be expressed
also in terms of GPDSRs  in the form of an integral equation
(\ref{Def-GDPSR}) or in terms of FESR-like ones
(\ref{Def-FEGDPSR}).

We now apply our results to the DVCS case, in particular,
to the dominant CFF  ${\cal H}$:
\begin{equation}
\vartheta=1\,, \qquad {\cal H}(\xi,t,{\cal Q}^2)\equiv {\cal
H}(\xi,\vartheta=1,t,Q^2)\,, \quad
{\cal C}(\xi,t,{\cal Q}^2)\equiv {\cal C}(\xi,\vartheta=1,t,Q^2) \, .
\end{equation}
Here we introduce the virtuality of the photon ${\cal Q}^2 \approx 2
Q^2$. Everything that was stated regarding the measurement of CFFs
remains valid in DVCS kinematics. It is often easier to measure
the imaginary part (via single spin asymmetries or polarized cross
sections) than the real part of the amplitude. Hence, one might
use the available experimental findings for the real part to
reveal the imaginary part of the CFF  outside of the accessible
phase space.

We introduce two phenomenological parameters $ \underline{\eta}$
and $ \overline{\eta}$ to divide the $\xi$ interval into the small
($\xi \le \underline{\eta}$), valence  ($\underline{\eta} \le \xi
\le \overline{\eta}$), and large ($ \overline{\eta} \le \xi$)
region. We assume that the  CFF can be measured in the valence
region $\underline{\eta} \le \xi \le \overline{\eta}$. The
dispersion integral (\ref{Def-DisRel}) might then be used to
constrain the  imaginary part in the large $\xi'$ region
$\overline{\eta} \le \xi'$
\begin{eqnarray}
\label{Exp-SumRul} \int_{\overline{\eta}}^{1}\!d\xi^\prime
\frac{2\xi^\prime }{\xi^2-{\xi^\prime}^2}
 \left[\Im{\rm m}{\cal H}(\xi^\prime,t,{\cal Q}^2)-\Im{\rm m}{\cal H}(\xi,t,{\cal Q}^2) \right]
= \pi {\cal S}^{\rm exp}(\xi,t,{\cal
Q}^2|\overline{\eta},\underline{\eta}) + \pi {\cal S}^{\rm
mod}(\xi,t,{\cal Q}^2|\underline{\eta}) \,,
\end{eqnarray}
from the experimental knowledge of the region $\underline{\eta}
\le \xi' \le \overline{\eta}$:
\begin{eqnarray}
\label{Exp-SumRul-Exp} {\cal S}^{\rm exp}(\xi,t,{\cal
Q}|\overline{\eta},\underline{\eta}) &\!\!\!=\!\!\!& \Re{\rm e}
{\cal H}(\xi,t,{\cal Q}^2) -  \frac{1}{\pi}
\ln\!\left(\frac{\xi^2}{1-\xi^2}\! \right)\,
 \Im{\rm m}{\cal H}(\xi,t,{\cal Q}^2)
\\
 &&\!\!\!+\frac{1}{\pi} \int_{\underline{\eta}}^{\overline{\eta}}\!
d\xi^\prime \frac{2\xi^\prime }{{\xi^\prime}^2-\xi^2}
\left[\Im{\rm m}{\cal H}(\xi^\prime,t,{\cal Q}^2)-\Im{\rm m}{\cal
H}(\xi,t,{\cal Q}^2)\right]\,, \nonumber
\end{eqnarray}
the model estimate for the subtraction constant, and the
extrapolation into the small $\xi'$ region $\xi'  \le
\underline{\eta}$:
\begin{eqnarray}
\label{Exp-SumRul-Mod} {\cal S}^{\rm mod}(\xi,t,{\cal
Q}|\underline{\eta}) = {\cal C}(t,{\cal Q}^2) +\frac{1}{\pi}
\int_0^{\underline{\eta}}\! d\xi^\prime \frac{2\xi^\prime
}{{\xi^\prime}^2-\xi^2}  \left[\Im{\rm m}{\cal
H}(\xi^\prime,t,{\cal Q}^2)-\Im{\rm m}{\cal H}(\xi,t,{\cal
Q}^2)\right] \,.
\end{eqnarray}
The subtraction constant is known to some extent, see, e.g.,
Refs.~\cite{GoePolVan01,Goeetal07,Hagetal07}, and one might
develop some understanding how  Regge phenomenology  should be
employed for off-shell processes. In evaluating the amount arising
from the very small $\xi$ region one can use the DVCS measurements
of the H1/ZEUS collaborations
\cite{Adletal01,Chekanov:2003ya,Aktas:2005ty,Aaretal07}. Hence,
the model dependent part is therefore somewhat under control. The
large $\xi$ region which one reveals in this way is certainly
challenging to measure in experiment. As one realizes, even
knowledge of the CFF at one value of $\xi$, i.e.,
$\underline{\eta}=\overline{\eta} =\xi$, leads to a highly
non-trivial constraint. This sum rule (\ref{Exp-SumRul}) (with $\xi
\le \overline{\eta}$) can again be considered  as a family of
FESR. To see this,  one simply expands the integral kernel with
respect to $(\xi/\xi^\prime)^2$. We also note that the real part
of the CFF  ${\cal H}$ can be measured in the crossed channel
$\gamma^\ast \gamma^{(\ast)} \to N \bar{N}$ (or $ N \bar{N} \to
l^+l^- \gamma$)  and then obtained by analytic continuation.

The dispersion integral and crossing symmetry allows for revealing the
CFF as a physical quantity. The microscopic view on this quantity
is given by perturbation theory, providing us, to LO accuracy, with a
partonic interpretation in terms of the GPD on the cross-over
trajectory:
\begin{eqnarray}
\label{Mea-DVCS-1} H^-(x,x,t,{\cal Q}^2) \stackrel{\rm LO}{=}
\frac{1}{\pi}\,\Im{\rm m}{\cal H}(\xi=x,t,{\cal Q}^2) \quad
\mbox{for} \quad \underline{\eta} \le x \le \overline{\eta}\,.
\end{eqnarray}
To that accuracy we write the physical sum rule (\ref{Exp-SumRul})
as a GPD one:
\begin{eqnarray}
\label{Exp-SumRul-1} &&\phantom{+} \int_{\overline{\eta}}^{1}\!dx
\frac{2x }{\xi^2-x^2}
 \left[ {H^-}(x,x,t,{\cal Q}^2) -H^-(\xi,\xi,t,{\cal Q}^2) \right]
 -{ C}(t,{\cal Q}^2)
 \\
 &&+
 \int^{\underline{\eta}}_0\!dx \frac{2x }{\xi^2-x^2}
 \left[ {H^-}(x,x,t,{\cal Q}^2) -H^-(\xi,\xi,t,{\cal Q}^2) \right]
 \stackrel{\rm LO}{=}
{\cal S}^{\rm exp}(\xi,t,{\cal
Q}|\overline{\eta},\underline{\eta})
\quad  \mbox{for} \quad  \underline{\eta}  \le \xi   \le
\overline{\eta} \,. \nonumber
\end{eqnarray}
It clearly states that in an experiment the GPD can be
accessed only on the cross-over trajectory.

The information which can be revealed away from the cross-over
trajectory is governed by the GPDSR family (\ref{Def-GDPSR}),
which we write in the form of an integral equation
\begin{eqnarray}
\label{Def-GDPSR-H} {\rm PV}\int_0^{1}\! dx\;
\frac{2x}{\eta^2-x^2} H^-(x,\eta,t,{\cal Q}^2) = {\rm PV}
\int_0^{1}\! dx\; \frac{2x}{\eta^2-x^2} H^-(x,x,t,{\cal Q}^2) -
C(t,{\cal Q}^2) \,.
\end{eqnarray}
The r.h.s.~of this integral equation is
considered as known and can be equated in an
ideal experiment with the measured real part of
the CFF (including the subtraction constant).
The l.h.s.~is nothing more than the real part of
the LO result (\ref{Def-AmpLO}). In addition, we
have the `boundary condition' that
$H^-(x,x,t,Q^2)$ is equal to the imaginary part
of the CFF. Knowing the imaginary and real part of
$\cal H$ in an ideal experiment, one would be
able to measure the subtraction constant.
However, the deconvolution problem cannot be
solved%
\footnote{We recall that the scale dependence provides
an additional handle on the deconvolution
problem.
}
within the family of GPDSRs (\ref{Def-GDPSR-H}).

The invisible terms in an ideal DVCS experiment are the zero
modes,
\begin{eqnarray}
\label{Def-GDPSR-H-0} \int_0^{1}\! dx\; \frac{2x}{\eta^2-x^2}
\delta H(x,\eta,t,{\cal Q}^2)
\equiv 0\,,
\end{eqnarray}
in the integral equation (\ref{Def-GDPSR-H}) with the boundary
conditions
\begin{eqnarray}
\delta H(x,\eta=x,t,{\cal Q}^2) =0\,, \quad  \lim_{\eta\to 0}
\delta H(x,\eta,t,{\cal Q}^2) = x^{\underline{\epsilon}}\,, \quad
\lim_{\eta\to 1} \delta H(x,\eta,t,{\cal Q}^2)
=(1-x)^{\overline{\epsilon}}\,,
\end{eqnarray}
where
$\left\{\underline{\epsilon},\overline{\epsilon}\right\}
>0$. Due to the duality properties, governed by
Lorentz covariance, it is clear that a priori any given function
in the outer or central region which satisfies the boundary
conditions defines a zero mode. A simple example for fixed ${\cal
Q}^2$ is
\begin{eqnarray}
\delta H^{\rm toy}(x,\eta,t) =   \frac{(\eta^2 - x^2)}{|x|}
\frac{d}{d x}  f(|x|,t)
\end{eqnarray}
with appropriate behavior of $f(x,t)$ at the boundaries $x=0$ and
$x=1$. Hence, the experimental data (at least for a fixed scale) can be
described by numerous GPD models. In our toy example, the zero
mode might contribute to the forward case and might be visible for
$t=0$ in DIS. An example of a term visible in DVCS, but invisible
in forward kinematics, is given in Ref.~\cite{GuiMor07}. Also,
for GPD $E^-$, where phenomenological constraints for the forward
limit do not exist, any GPD model which fits the data is as good
as any other, obtained by a `gauge' transformation. However, the
zero mode might contribute to the first moment:
\begin{eqnarray}
E^{-{\rm fit}}(x ,\eta ,t,{\cal Q}^2) \to E^{-{\rm fit}}(x ,\eta
,t,{\cal Q}^2) + \delta E(x ,\eta ,t,{\cal Q}^2)\,, \quad
\int_{-1}^1\!dx\, x\, \delta E(x ,\eta ,t,{\cal Q}^2) \neq 0\,.
\end{eqnarray}
Obviously, different GPD models which are able to describe the
experimental data can provide different values for the desired
first moment, which enters the quark angular momentum sum rule
\cite{Ji96}.

Although the GPDSR family (\ref{Def-GDPSR-H}) states that the
central and the outer GPD region are related via the GPD on the
cross-over trajectory, some information might be lost in DVCS
and/or for forward kinematics. For a more geometrical
interpretation we refer to the discussion in the Introduction and
Fig.~\ref{Fig-GPDans}(b). Only in the case that for GPDs the
holographic principle applies, i.e., all information which is
contained in the GPD can be obtained from its value on the
cross-over trajectory, the whole GPD can be revealed from an
(ideal) measurement. Nevertheless, the GPD sum rules
(\ref{Def-GDPSR-H}) provide us with valuable information on the
$\eta$-dependence of the GPD.

This can be seen if we decompose the integral into contributions
from the central and outer regions. Neglecting a possible
subtraction constant and assuming that the Regge behavior is
$\eta$-independent,
\begin{eqnarray}
\label{Sum-Rul-DVCS1} \int_0^{\eta}\! dx\; \frac{2 x}{\eta^2-x^2}
\left[H^-(x,x,t)-H^-(x,\eta,t)\right]= \int_\eta^{1}\! dx\;
\frac{2x}{x^2-\eta^2} \left[H^-(x,x,t)-H^-(x,\eta,t)\right]\,,
\end{eqnarray}
we find that in the limit $\eta\to 0$ the l.h.s.~vanishes and the
r.h.s.~must therefore satisfy an inverse moment sum rule. Taking
now the limit $\eta\to 1$ and using the fact that the GPD is vanishing at
the end point $x=1$, we find another sum rule. These sum rules
read
\begin{eqnarray}
\label{Sum-Rules-GPD} \int_0^{1}\! dx\; \frac{1}{x}
\left[H^-(x,x,t)-H^-(x,0,t)\right]=0 \,, \quad \int_0^{1}\! dx\;
\frac{x}{1-x^2} \left[H^-(x,x,t)-H^-(x,1,t)\right]= 0\,.
\end{eqnarray}
In the left sum rule (i.e., $\eta\to 0$) the GPD on the cross-over
trajectory is dictated by the physical content in the outer
region, while  in the right sum rule (i.e., $\eta\to 1$) it is
governed by the one in the central region. This suggests that for
small $\eta$ we shall naturally have the partonic $s$-channel point
of view, while for large $\eta$ the $t$-channel point of view in
terms of meson-like exchanges is more appropriate \cite{PolShu02};
see left and right sketch in the lower part of the
Fig.~\ref{Fig-Dua}(b), respectively. To match both regions one
might sum up the $t$-channel exchanges, i.e., reggeize them. If
one would like to incorporate phenomenological knowledge of
hadronic physics in GPD modelling,  one might suppose that for
small $x$  and $\eta$  one understands better the hadronic
$t$-channel view in terms of mesonic Regge exchanges
\cite{KumMuePas07}, while for large $\eta$ and $x$ the $s$-channel
partial wave expansion in terms of few baryonic resonances is more
useful \cite{CloZha02}; see right and left   sketch in the upper
part of the Fig.~\ref{Fig-Dua}(b), respectively. Hence, GPD
modelling from phenomenological input should be related as
indicated across the four sketches in the Fig.~\ref{Fig-Dua}(b)
rather then vertically. Finally, we note that  the left equation in
Eq.~(\ref{Sum-Rules-GPD}) is given in its  more general form in
Eq.~(\ref{Sum-Rul-DVCS-0}).

Let us also discuss the problem of relating the GPD at the
cross-over trajectory  with the zero skewness GPD from the point
of view of Mellin moments. The subtraction constant that enters
the DVCS process is, according to Eq.~(\ref{eq:CsumFnn1}), given by
\begin{eqnarray}
C(t,{\cal Q}^2) = - 2\sum_{\substack{j=1 \\ {\rm odd}}}^\infty
h^{(j+1)}_j(t,{\cal Q}^2) \, . \label{eq:CsumFnn1theta1}
\end{eqnarray}
A model dependent estimate of the constant for $t=0$ was provided
in $\chi$QSM \cite{GoePolVan01}. The $\chi$QSM model
evaluation of the first term in the series
(\ref{eq:CsumFnn1theta1}), given in Ref.~\cite{Goeetal07}, is
compatible with lattice measurements \cite{Hagetal07}. This allows one
to estimate the size of the  constant $C(t,{\cal Q}^2)$, to judge on the
convergence of the series (\ref{eq:CsumFnn1theta1}), and to
estimate its $t$-dependence. According to Eq.~(\ref{Def-Melmomxx}),
the Mellin moments of the GPD at the cross-over trajectory
are given by the series
\begin{eqnarray}
\int_{0}^1\!d x\, x^j H^-(x,x,t,{\cal Q}^2)  = \sum_{\substack{n=0 \\ {\rm
even}}}^\infty h_{j+n}^{(n)}(t,{\cal Q}^2)\,.
\end{eqnarray}
The first term with $n=0$ in this series is the $j^{ th}$ Mellin moment of the GPD at $\eta=0$.

Instead of using the number of total derivatives $n$ as a label for
the degrees of freedom, one might prefer the angular momentum $J$
of the $t$-channel SO(3) partial wave expansion, which is a physical
degree of freedom. If we apply
crossing symmetry to  the GPDSR family (\ref{Def-GDPSR-H}), for
details see, e.g., Ref.~\cite{MueSch05}, and set $\eta^2 \approx
1/\cos^2\theta$, where $\theta$ is the scattering angle in the
$t$-channel center-of-mass frame, we obtain the LO approximation
of the corresponding $\gamma \gamma^{\ast}\to p \overline{p}$
amplitude  in terms of the generalized distribution amplitudes
$H^{(t)-}(z,\cos\theta,s,{\cal Q}^2)$
\cite{MueRobGeyDitHor94,DieGouPirTer98}:
\begin{eqnarray}
\label{Def-GDPSR-H-1} \int_0^{1}\! dz\;  \frac{2z\,
}{1-z^2} H^{(t)-}(z,\cos\theta,s,{\cal Q}^2) = \int_0^{1}\!
dx\; \frac{2x\, \cos^2\theta}{1-x^2\, \cos^2\theta} H^-(x,x,t,{\cal Q}^2)
- C(t,{\cal Q}^2)\Big|_{t\to s} \,.
\end{eqnarray}
In the CFF combination ${\cal H} + t\, {\cal E}/4 M^2_N $ Legendre
polynomials $P_J(\cos\theta)$ with angular momentum
$J=0,2,4,\cdots$ are the proper partial waves
\cite{Die03a}. The corresponding  partial wave amplitudes
$a_J(s,{\cal Q}^2)$ can be obtained from the orthogonality relation for
Legendre polynomials. In other words, one multiplies the l.h.s.~or r.h.s.~of, e.g.,
Eq.~(\ref{Def-GDPSR-H-1}) with the partial waves and integrates over
$Z=\cos\theta$. It is then clear from
Eq.~(\ref{Def-GDPSR-H-1}) that the subtraction constant $C(s,{\cal
Q}^2)$ contributes only to the $J=0$  partial wave amplitude. For
$J \ge 2$ the r.h.s.~of (\ref{Def-GDPSR-H-1}) tells us that the
partial wave amplitudes can be obtained from the Froissart--Gribov
projection by analytic continuation in $t$:
\begin{eqnarray}
a_J(s,{\cal Q}^2) \stackrel{\rm LO}{=} 2\int_{0}^1\! dx\,
\frac{{\cal Q}_J(1/x)}{x^2} \, \left[H^-(x,x,t,{\cal Q}^2) +
\frac{t}{4 M^2_N} E^-(x,x,t,{\cal Q}^2)\right]\Big|_{t\to s},
\label{Def-aJ}
\end{eqnarray}
where ${\cal Q}_J(z)$ are the Legendre functions of the second
kind. The representation (\ref{Def-aJ}) is suited for the analytic
continuation in $J$ within $ \alpha(t) < \Re{\rm e}{J} $, where
$\alpha(t)$ is the leading Regge trajectory. The partial wave
amplitudes are now, via Eq.~(\ref{Def-aJ}), represented in terms of
partonic quantities, namely, by the GPD on its cross-over
trajectory. We might also express them as a series of common Mellin
moments, labelled by the spin and the number of total derivatives,
or by the moments with definite conformal spin \cite{KumMuePas07}.

Obviously, the perturbative approach `predicts' the partial wave
amplitudes in terms of GPDs. Since the GPDs are unknown, it is more
appropriate to employ the reverse logic. As far as the partial wave
amplitudes are known they provide a strong constraint on the
GPDs. A power-like parameterization of the  GPDs at its cross-over
trajectory at small $x$, i.e., $x^{-\alpha(t)}$, corresponds to a
pole at $J=\alpha(t)$ in the SO(3) partial wave amplitudes and
makes direct contact to Regge phenomenology, extended to off-shell
kinematics. For integral values of $J\ge 2$ we can read
Eq.~(\ref{Def-aJ}) as a sum rule. The case $J=0$ deserves special
attention. It can be unproblematically evaluated from
Eq.~(\ref{Def-GDPSR-H-1}). Naively setting $J=0$ in
Eq.~(\ref{Def-aJ}) provides a divergent integral as the natural
outcome of our standard derivation. We should require that both
$J=0$ results are equal, which finally would provide a {\em
definition} of the inverse moment for the GPD at its cross-over
trajectory. Such an inverse moment is then simply given by the
subtraction constant $C(t,{\cal Q}^2)$.

The GPDSRs, such as in Eq.~(\ref{Def-GDPSR-H}), might be applied in various
manners and we have presented here only a few illustrative examples. Finally,
we would like to point out that they also  make contact to the work of
Refs.~\cite{PolShu02,Pol07,Pol07a}. Reducing the GPDSRs
(\ref{Def-FEGDPSR}) to the DVCS case, i.e., $\vartheta=1$, we find
in the limit $\upsilon \to 0$:
\begin{eqnarray}
\label{Def-FEGDPSR-DVCS}
\int^1_0\! dx\;
\left[ x^{-m-2} F^{\mp}(x,x ,t,Q^2)
- \sum_{\substack{n=0 \\ {\rm even}}}^{m+1} \frac{x^{n-m-2}}{n!}
\frac{d^{n}}{d\eta^{n}}F^{\mp}(x,\eta,t,Q^2)\Big|_{\eta=0} \right]
=
\lim_{\upsilon\to 0}\delta S^{\mp}_{m+1}(\upsilon,t,{\cal Q}^2)\,,
\end{eqnarray}
where $S^{\mp}_{m+1}(\upsilon,t,{\cal Q}^2)=S^{\mp}_{m+1}(\upsilon,\vartheta=1,t,Q^2)$ can
be read off from Eq.~(\ref{Def-FEGDPSR-S}). If we now view all terms as analytic functions
of $m$, we can invert this specific GPDSR (\ref{Def-FEGDPSR-DVCS}), by an inverse Mellin
transform to express the GPD at its cross-over trajectory in terms of `low-', i.e., second
term in the square brackets of Eq.~(\ref{Def-FEGDPSR-DVCS}), and  `high-energy' constants,
i.e., the terms of $S^{\mp}_{m+1}$.  This can then also be converted in a series of
forward-like parton distributions $q_\nu(x,t,{\cal Q}^2)$ weighted with $x^{\nu}$. The
expansion of the GPD on its cross-over trajectory  in this forward-like parton
distributions can be written, e.g., for $H^-$, as%
\footnote{One might find this result directly from the `Taylor
expansion' of $H^-(x,\eta,t,{\cal Q}^2)$ in the vicinity  of
$\eta=0$ with $x=\eta$. However, the expansion coefficients are
generalized functions, whose property (47) ensures polynomiality. These
forward-like parton distributions can be straightforwardly evaluated from DDs
and the expansion (\ref{Rep-Dual}) can be studied,
e.g., within a RDDA. }
\begin{eqnarray}
H^-(x,x,t,{\cal Q}^2) = \sum_{\substack{\nu=0 \\  {\rm even}}}^\infty\,
x^\nu\, q_\nu(x,t,{\cal Q}^2)\,\quad\mbox{with}\quad  q_0(x,t,{\cal Q}^2)\equiv H^-(x,\eta=0,t,{\cal Q}^2)\,. \label{Rep-Dual}
\end{eqnarray}
Here the forward-like parton functions know about both the `low-'
and `high-energy' part and should be viewed as generalized
functions in the mathematical sense. The polynomiality condition
translates into
\begin{eqnarray}
\int_0^1\! dx\, x^{m-1} \, q_n(x,t,{\cal Q}^2) =0 \quad\mbox{for}\quad   2\le m < n\,,
\label{Con-Pol}
\end{eqnarray}
where $m$ and $n$ are even. The decomposition of these
forward-like parton functions in forward-like conformal or `dual'
parton functions has been discussed in the small $\eta$
expansion in Ref.~\cite{Sem08}. Generally, it should be given by
an integral transformation,
cf.~Refs.~\cite{Shu99,Nor00,PolShu02,Pol07,Pol07a}. We emphasize
that these forward-like functions, e.g., given by explicit
expressions in Ref.~\cite{Pol07a}, can be interpreted in terms of
GPDSRs (\ref{Def-FEGDPSR-DVCS}). The DVCS amplitude to LO in the
`dual' parameterization is then finally represented as a sum of
functions that are associated with the `low-' and `high-energy'
content, where the duality scale $\upsilon$ itself is taken to be
zero; see Ref.~\cite{Pol07a}. This step generates the
$\delta$-functions (including derivatives) which appear in the
`dual' parameterization. We recall that in the series
(\ref{Rep-Dual}) all forward-like parton distributions contribute
to the small $x$ region and that at least the first two terms are
needed to describe DVCS to LO accuracy in the small $x_{\rm Bj}$
region.

\subsection{Revealing GPDs from experimental measurements}
\label{SubSec-RevGPD}

We address now the question: ``What do we learn from  a  DVCS
measurement?''. To have a partonic interpretation one is mostly
interested in the GPD at $\eta=0$. The extrapolation of
$F^{\mp}(x,x,t,{\cal Q}^2)$ to $F^{\mp}(x,0,t,{\cal Q}^2)$ is a
model dependent procedure, which can be parameterized by the
skewness function $S(x,t,{\cal Q}^2|F^{\mp})$ defined by
\begin{equation}
\label{Def-skePar}
 F^{\mp}(x,x,t,{\cal Q}^2) = \Big[1+ S(x,t,{\cal
Q}^2|F^{\mp})\Big] F^{\mp}(x,\eta=0,t,{\cal Q}^2)\,.
\end{equation}
Note that we included a $t$-dependence in the skewness function, since
it is expected that this dependence at large $x$ is rather
different from that of the GPD at $\eta=0$. The GPDSR family
(\ref{Def-GDPSR}) provides us in the limit $\xi \to 0$ with a link
between the measured GPD at $\eta=x$ and $\eta=0$ and therefore a
constraint for the skewness function. Since the limit might be
intricate, we first have to regularize:
\begin{eqnarray}
\label{Sum-Rul-DVCS-0} \int_{(0)}^{1}\! dx\; \frac{1}{x}
S(x,t,{\cal Q}^2|F^-) F^{-}(x,\eta=0,t,{\cal Q}^2) =\frac{1}{2}{
C}_F(t,{\cal Q}^2) \,. \qquad
\end{eqnarray}

One might consider the $F^\mp(x,x,t,{\cal Q}^2)$ as new functions,
which are in an unknown manner connected to the GPD at $\eta=0$.
As emphasized above, the analytic regularization procedure might
be reinterpreted in terms of partonic FESR. The constraint
(\ref{Sum-Rul-DVCS-0}) on the skewness effect is rather weak; only
the inverse moment is constrained, and so there are infinite
degrees of freedom left, e.g., in the series
(\ref{eq:CsumFnn1theta1}). We might parameterize the skewness
effect by relative deviation factors:
\begin{eqnarray}
\label{Cal-del} \delta_j(t,\mu^2 | F^\mp)  \equiv  \sum_{\substack{n=2 \\ {\rm
even}}}^\infty \frac{f_{j+n}^{(n)}(t,\mu^2)}{ {f}_j(t,\mu^2)} =
 \frac{\int_{0}^1\! dx\, x^j
S(x,t,\mu^2|F^\mp)F^\mp(x,\eta=0,t,\mu^2)}{\int_{0}^1\! dx\, x^j
F^\mp(x,\eta=0, t,\mu^2)}\,,
\end{eqnarray}
with, as before,  $j$ being odd (even) for the $-$ ($+$)
superscripts. The Mellin moments of CFFs are then given as
a product,
\begin{eqnarray}
\int_{0}^1\!d\xi\, \xi^j \Im{\rm m}{\cal F}(\xi,t,{\cal Q}^2)
\stackrel{\rm LO}{=} \pi \int_{0}^1\!d\xi\,\xi^j
{F}^\mp(x=\xi,\xi,t,{\cal Q}^2)  =  \pi f_j(t,{\cal Q}^2) \left[1+
\delta_j(t,{\cal Q}^2 | F^\mp)\right]\,,
\end{eqnarray}
of the desired Mellin moment $f_j(t,{\cal Q}^2)$, surviving the
forward limit, times the skewness effect.

For $t=0$ the Mellin moments $h_j(t=0,{\cal Q}^2)$ [$\widetilde
h_j(t=0,{\cal Q}^2)$] are those of the PDFs and are constrained by
phenomenological input. The lowest Mellin moments $f_{j=0}(t,{\cal
Q}^2)$ are partonic form factors, related to observables. The
deviation factors are built by a series (\ref{Cal-del}) and the
present state of the art in lattice measurements, to the 
best of our knowledge, allows for the evaluation
of the first term in this series for $j=0$, i.e.,
the quantity $h^{(2)}_2$. As long as the
forward limit is known (to some extent) a DVCS measurement of,
e.g., $\cal H$ allows us to pin down the skewness effect for this
quantity. In the case one wants to extract, for instance, the
quark angular momentum in an ideal DVCS experiment via the anomalous
gravitomagnetic  moment \cite{Ji96},
\begin{eqnarray}
B({\cal Q}^2) \equiv e_1(t=0,{\cal Q}^2)\stackrel{\rm LO}{=} \frac{1}{\pi} \frac{
\int_{0}^1\! d\xi\, \xi\, \lim_{t\to 0}\Im{\rm m} {\cal E}(\xi, t,{\cal
Q}^2)}{1+\delta_1(t=0,{\cal Q}^2 |E^-)}\,,
\end{eqnarray}
one certainly needs to know the skewness effect (at $t=0$), which might be
different from that appearing in $\cal H$.

The conclusion is obvious. One must use additional information to
reveal the quantities one likes to  pin down. Lattice measurements
can give only a hint on the size of the deviation factors, and this we
consider very valuable. Model evaluations, e.g., within $\chi$QSM,
can at least  give estimates for the size of deviation factors. As
in any of such model calculation, the matching of collective with
current quark degrees of freedom induces uncertainties.
Nevertheless, it is very important to quantify these deviation
factors on a generic level. Further experimental constraints on
them, available in the small $x_{\rm Bj}$ region, arise from the
logarithmical scaling violation, governed by evolution. A
discussion of revealing the quark angular momentum from the present
DVCS data should be given elsewhere. We would only like to
note here that GPD model dependent constraints on the quark
angular momentum require understanding of the skewness effect.
A minimal basic requirement for such a constraint is that the
utilized GPD model describes all available DVCS data, including
the presently challenging data on unpolarized cross section.

According to the above discussion, one strategy for extracting GPD
information to LO accuracy might consist of the following steps:
(suppose that we consider $H$ or $\widetilde H$)
\begin{itemize}
\item
model the $t$-dependency  of the GPD at $\eta=0$,
\item
parameterize the skewness function, defined in Eq.~(\ref{Def-skePar}), and
\item
fit the parameters to the measured observables (CFFs) by means of
Eqs.~(\ref{Mea-DVCS-1}) and (\ref{Exp-SumRul-1}).
\end{itemize}
Alternatively, in the third step one might use the sum rule
(\ref{Exp-SumRul}) to first pin down the imaginary part in the
whole region from knowledge of the real part and then reveal 
from Eq.~(\ref{Mea-DVCS-1}) the GPD at the cross-over trajectory.

Certainly, in the first step it is crucial to have a realistic
model for the interplay of the $x$- and $t$-dependence of the $\eta=0$
GPD. Proposals for the functional form of the GPD at $\eta=0$ are
given in Ref.~\cite{DieFelJakKro04,GuiPolRadVan04,AhmHonLiuTan06}.
We would like to suggest another empirical recipe that allows one to
take advantage of theoretical/phenomenological knowledge on the
generic level. Moreover, it has a power-like behavior at large
$-t$ and can in principle also be used for positive values of $t$.
Inspired by the simplicity of the Veneziano model \cite{Ven68} for
on-shell amplitudes, we found it simple to set up GPD models at
$\eta=0$ in Mellin space by taking ratios of $\Gamma$ functions,
e.g., for the valence $d$-quark contribution:
\begin{eqnarray}
h_j^{d_{\rm val}} = \frac{\Gamma (1+j -\alpha(t)) \Gamma (1+j+P-\alpha)
\Gamma (2-\alpha +\beta )}
{\Gamma (1-\alpha) \Gamma (1+j+P-\alpha(t)) \Gamma (2+j-\alpha +\beta)}\,.
\end{eqnarray}
The generic  values of the parameters arise from counting rules
($\beta=3$, $P=2$), Regge phenomenology ($\alpha(t)=1/2 + t/{\rm
GeV}^2$), and for $u$-quarks we employ in addition SU(6) symmetry
arguments, i.e., $h_j^{u_{\rm val}}/h_j^{d_{\rm val}}=5$ for
$j\to\infty$. These parameters are then understood as effective
ones and are slightly tuned to phenomenological PDF
parameterizations and form factor measurements. Lattice results
\cite{Hagetal07} can be quantitatively  reproduced within our
ansatz if we adopt the found parameterization of the Regge
trajectory $\alpha(t)$ to the large meson mass  lattice
world.%
\footnote{We do not share the wide-spread belief, e.g., in
Ref.~\cite{AhmHonLiuTan07}, that present lattice data should be
used as quantitative GPD constraints.}
The guidance for the
parameterization of the skewness function $S(x,t,{\cal
Q}^2=2\,{\rm GeV}^2|H^-)$ comes from the functional form of the
solution for the positivity constraints \cite{Pob03}, the expected
large $x$ behavior for GPDs at $\eta=x$ \cite{Yua03}, and model
results \cite{HwaMue07}. The skewness function satisfies the sum
rule (\ref{Sum-Rul-DVCS-0}), where the $t$-dependence of the
subtraction constant has been taken from the first coefficient of
the $D$-term from the $\chi$QSM result \cite{Goeetal07}. The small
$x$ behavior of the GPD $H$ is taken from our LO fits to the
H1/ZEUS data \cite{KumMuePasSch08}.  The modelling will be
described in detail somewhere else.

\begin{figure}[t]
\centerline{\includegraphics[scale=0.95]{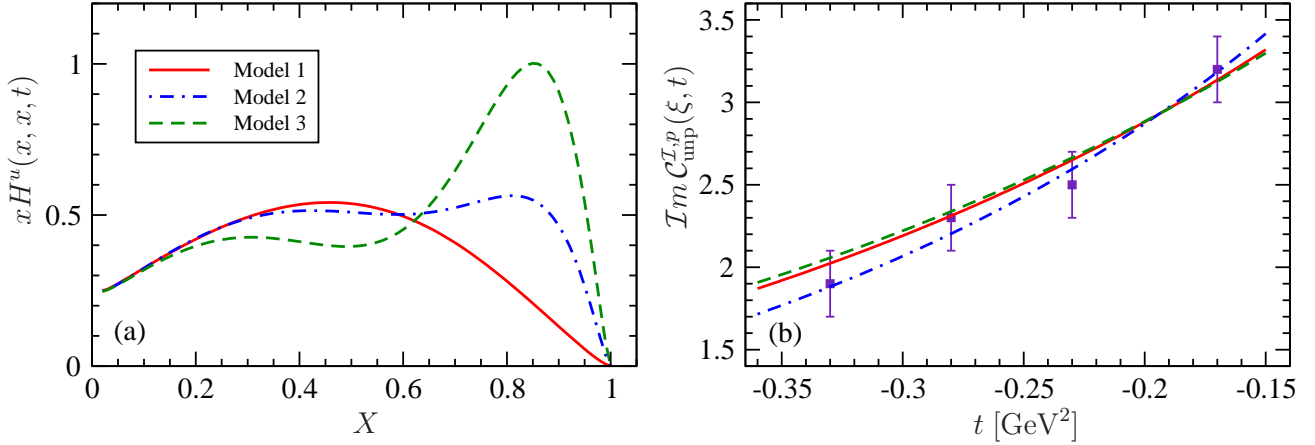}}
\caption{\label{ComJLAB} Panel (a) shows various models for the
dominant $u$-quark GPD $H$ at $\eta=x$ versus the Bjorken-like
variable $X=2x/(1+x)$ for fixed $t=-0.25\, {\rm GeV}^2$. In panel
(b) we confront these models with JLab/Hall A measurements of the
observable $\Im{\rm m}\,{\cal C}^{{\cal I},p}_{\rm unp}$, defined
in Eq.~(\ref{Def-CIntunp}), for DVCS off unpolarized proton
\cite{Cametal06} for $x_{\rm Bj}=0.36$ and ${\cal Q}^2 =1.9\, {\rm
GeV}^2$. }
\end{figure}
In Fig.~\ref{ComJLAB}(a) we show various versions of the model for
the dominant $u$-quark GPD $H$. Only the solid curve is rather
similar to the model versions of Ref.~\cite{BelMueKir01}; the
two other ones belong to a new class of GPD models.   As argued in
Sect.~\ref{SubSecLes}, they are enhanced in the large
$X=2x/(1+x)$ region. Note that such  a functional form resembles the shape
of the energy spectrum function of an exclusive process, if 1/X is
viewed as a rescaled energy (perhaps one would even expect a resonance
structure before the `continuum' starts). The $u$-quark GPD $H$ is
the dominant one in DVCS off an unpolarized proton target and it
might be revealed from the azimuthal dependence of the
interference term. The quantity which determines the magnitude of
the dominant first harmonic in the interference term \cite{BelMueKir01},
\begin{eqnarray}
\label{Def-CIntunp} {\cal C}^{\cal I}_{\rm unp} = F_1(t) {\cal
H}(\xi,t,{\cal Q}^2) + \xi
\left[F_1(t)+F_2(t)\right]\widetilde{\cal H}(\xi,t,{\cal Q}^2) -
\frac{t}{4 M^2} F_2(t) {\cal E}(\xi,t,{\cal
Q}^2)\big|_{\xi=\frac{x_{\rm Bj}}{2-x_{\rm Bj}}}\,,
\end{eqnarray}
is shown in Fig.~\ref{ComJLAB}(b) for  JLab/Hall A kinematics:
$$
x_{\rm Bj}=0.36\,, \;\; {\cal Q}^2 =1.9\, {\rm GeV}^2\,,\;\;
\mbox{and}\;\; 0.15\, {\rm GeV}^2 \le -t \le 0.35\, {\rm
GeV}^2\,.$$ The relative contribution of the terms involving CFFs
$\widetilde {\cal H}$ and ${\cal E}$ we estimate to be on the
level of 10\%--40\% and up to $\pm 10\%$, respectively, compared
to the dominant ${\cal H}$ term. Instead of exploiting the sum
rule family (\ref{Exp-SumRul}), we confronted our GPD model with
experimental data in the old fashioned way. We varied the unknown
contributions and compensated for this by tuning the skewness function
for $H$. We  emphasize that any curve in Fig.~\ref{ComJLAB}(a)
that crosses one of the shown ones at $X=0.36$ defines a priori a
GPD model, which is not ruled out by the JLab/Hall A data in panel
(b). The GPD models are then selected  by confronting them with
the measurement for the real part for the observable
(\ref{Def-CIntunp}). We found that the extracted values for this
quantity are contaminated by  DVCS squared terms and so there is a
further contribution from the GPD $\widetilde E$. Taking all this
into account we had no problems to tune the model parameters to
the measured values for $t=-0.25\, {\rm GeV}^2$. However, we were
only able to get a qualitative agreement with the measured
$t$-dependence of this quantity. Whether this is an artifact of
our own GPD model restrictions or is due to the power
suppressed terms in the twist-three formalism of
Ref.~\cite{BelMueKir01} remains for the moment an open question.
This is in our opinion more a technicality than an issue, which,
of course, must be investigated.

\section{Outlook}
\label{Sec-Out}

We have considered CFFs as holomorphic functions and discussed the
phenomenological application of this property. To LO accuracy and
for fixed photon virtuality one can in a DVCS measurement access
the GPD only on its cross-over trajectory $\eta=x$. Combining
dispersion relation and perturbation theory allows for a derivation of
the partonic analogue of FESR, which relate the small and large $x$
behavior of the GPD. Altogether, this
sets up a new strategy for revealing GPDs from experimental DVCS
measurements, utilizing well-known techniques from the
phenomenology of on-shell hadronic processes. The proposed
alternative LO GPD approach has numerous advantages with respect
to the present ad hoc GPD model approach. The LO GPDs can be
considered as `super' GPDs that contain all perturbative and
non-perturbative corrections. If the assumed holomorphic
properties hold true,  the approach by definition cannot fail to
describe DVCS data for fixed photon virtuality. All discussions
about GPD models/ans\"atze and erroneous conclusions coming from their
success or failure are avoided from the very beginning. The goal
of this concept is to reveal GPDs at their cross-over trajectory
$\eta=x$ only from DVCS data and to obtain a generic understanding
of the skewness effect.

The partonic interpretation of the GPD to LO accuracy has been
given here only from the principal point of view, where  the
appropriate choice of scaling variables was not discussed. This
rather important problem is related to the minimization of not
well-known higher twist contributions, by appropriate variable
definitions. We recall that a partonic interpretation of the
`measured' GPD at the cross-over trajectory has already been
pointed out in Ref.~\cite{RalPir01}. We would like to specialize this
interpretation. In the infinite momentum frame one might label the
proton state by the center of plus-momentum, given as the expectation
value $\mbox{\boldmath $ b$}_\perp$  of the two generators for
transversal Galilean boost \cite{KogSop70}. The emission of a
parton with momentum fraction $X=2x/(1+x)$ leads to a change of
the center of plus-momentum $\mbox{\boldmath $ b$}_\perp^{\rm in}$
of the initial nucleon. Since for $\eta=x$ the outgoing spectator
system absorbs a quark with zero longitudinal momentum fraction,
its center of plus-momentum will not change and it coincides with
the one of the final state nucleon $\mbox{\boldmath $
b$}_\perp^{\rm spec}=\mbox{\boldmath $ b$}_\perp^{\rm out}$. The
difference $ \mbox{\boldmath $\delta b$}_\perp= \mbox{\boldmath $
b$}_\perp^{\rm in}-\mbox{\boldmath $ b$}_\perp^{\rm spec}$ can
now be viewed as the transversal distance of the struck parton from
the collective spectator system. One might now adopt the GPD
wave function overlap interpretation in the impact parameter space
\cite{Die02}, which tells that the Fourier conjugate variable of
$\mbox{\boldmath $\delta b$}_\perp$ is simply the transverse
momentum transfer $ \mbox{\boldmath $\Delta$}_\perp$, directly
observed in experiment.
In other words, following the procedure of
Ref.~\cite{RalPir01} gives us access to the transversal separation
distance of the struck quark and collective spectator system.%
\footnote{This interpretation has been also presented in Ref.~\cite{Bur07},
including considerations on the $t$-behavior of GPDs at large
$\eta=x$ as it arises also  from power-like wave function models
\cite{MukMusPauRad02,TibDetMil04,HwaMue07}.
} 
A probabilistic
GPD interpretation as for the zero skewness case, however, does
not hold here anymore.

Also, the skewness effect can be addressed from the experimental
side. In DVCS observables for an unpolarized nucleon target one might
first neglect the non-dominant CFFs $\widetilde {\cal H}$, ${\cal
E}$ and $\widetilde {\cal E}$ as well as  twist-three
contributions. This seems to be justified to a certain extent, at
least for HERMES kinematics. For the dominant quantity ${\cal H}$
the corresponding GPDs in the zero skewness case are well
constrained by PDFs and form factors. Guidance from model
calculations and lattice data might be employed with care, too, to
set up  field theory inspired GPD models for the zero skewness
case. The skewness function, which relates the GPD for zero
skewness  to the one on the cross-over trajectory
is constrained by GPD sum rules (GPDSR). It can then
directly be fitted to  experimental data.  The limitation of this
pragmatical approach for the collider experiments H1 and ZEUS is
also obvious: the perturbative prediction of the scale change, which
gives us an additional handle on the skewness effect, cannot be
implemented. Since the functional form of CFFs and GPDs
drastically simplify in this kinematics, experimental data can be
analyzed in the traditional manner within a flexible GPD
parameterization  in the small $x$ region.  This can be done in
any GPD representation; however, it has up to now only been
applied in the conformal approach.

We expect that a global LO analysis of present DVCS data can
already provide a generic understanding of the GPD at the
cross-over trajectory, in particular for the dominant GPD $H$.
Once this understanding is reached one can address the
deconvolution problem, i.e., to find a realistic parameterization
of GPD models. This is particularly important if in dedicated
experiments the CFFs $\cal E$ and $\widetilde{\cal E}$ are
measured and one wants to access the GPDs $E$ and $\widetilde
E$, where the former one gives access to the quark angular
momentum. Realistic and flexibly parameterized GPD models should
then open the road to the inclusion of radiative corrections and the
Wandzura-Wilczek approximation of twist-three GPDs, see, e.g.,
Ref.~\cite{AniPirTer00,KivPol00}, or to address quark-gluon-quark
GPDs \cite{BelMue00a}. If experimental data for virtual
electroproduction of mesons are analyzed within the GPD formalism,
it might be inevitable to go to next-to-leading order
accuracy. The formalism has been worked out for the twist-two sector
\cite{BelMue97a,BelMue98c,BelFreMue99,BelMue01a,IvaSzyKra04}.

In this paper it has clearly been spelled out that GPDs have dual
properties. If CFFs possess Regge behavior, this duality relation
is perfect and the GPDs contain the same information in the central
and outer region. This duality property is implemented in a GPDSR
family and can be expressed in terms of GPD moments, too.
Unfortunately,  these GPDSRs are not sufficient to find a unique
GPD model from (ideal) experimental DVCS measurements.  There are
two possible ways to get an handle on the remaining degrees of freedoms.

The first way relies on the common QCD tools:  phenomenological
knowledge, counting rules, model calculations, and lattice
simulations. Certainly, lattice simulations can only provide a
flash on the skewness effect, while model calculations suffer from
the matching problem of partonic and collective degrees of
freedom. A central theoretical task here is to understand the
small $x$ behavior of GPDs. It seems to be rather promising to
study the $t$-channel ladder, e.g., to proceed along the lines of
Refs.~\cite{FraFreGuzStr97} or \cite{ErmGreTro00}. We emphasize
that phenomenological knowledge can be implemented from the
$t$-channel point of view for small $x$ and also from the
$s$-channel point of view for large $x$, where both regions are tied
together by GPDSR. This suggests to look for dual GPD models in
which the meson-baryon duality conjecture, employed in the
phenomenology of hadronic on-shell processes, is implemented.
Crossing relations via the Froissart--Gribov  projection provide
an additional phenomenological handle.

The second way to address the problem is to develop a theoretical
understanding of whether there is some holographic principle that
tells us how the information about the GPD on the cross-over trajectory
is related to its values in the central (or outer)
region, which includes the $\eta=0$ case. This question arises
naturally if one relates the partonic GPD interpretation to the
$t$- and $s$-channel hadronic point of view. We recall that the
conjecture of duality originated in experimental findings about
hadronic on-shell processes and was extensively studied within
FESR; see, e.g., Ref.~\cite{Col77}. Thereby, the SO(3) partial
wave expansions in both the $t$- and $s$-channel were utilized in
the context of Regge phenomenology, which provides an empirical
description of the strong interaction phenomena. It was also
realized that duality phenomena could be partially explained
within the quark model. In  dual hadronic resonance models the
underlying symmetry of duality  is the SO(2,1) group. Finally, the
quark model was gauged providing us QCD, and  string theory was
born out of dual hadronic resonance models, i.e., out of the group
SO(2,1). Certainly, in QCD phenomenology quark-hadron duality, in
particular duality between the deep-inelastic and the resonance
kinematical regime (Bloom-Gilman duality) is an important
phenomenological concept; for a review see
Ref.~\cite{MelEntKep05}. Also conformal symmetry shows up in
various QCD corners
\cite{BroFriLepSac80,EfrRad80,Lip85,Lip93,Mue97a,BraKorMue03} and
integrability phenomena in the multi-color limit were discovered
in QCD \cite{Lip94,FadKor94,Kor94,BraDerKorMan99,Bel99} and in
supersymmetric Yang-Mill theories; see, e.g.,
Refs.~\cite{Bei04,BelKorMue05a} and references therein.  These
integrability phenomena play a key role in getting a handle on the
AdS/CFT conjecture \cite{Mal97,GubKlePol98,Wit98}. Initiated by
this conjecture, model dependent approaches to a supposed AdS/QCD
duality were developed \cite{BroTer03,BroTer05,GriRad07}, see also
Refs.~\cite{PolStr01,BroPolStrTan07}, and its application to
high-energy physics is discussed, e.g., in Ref.~\cite{HofMal08}. {From} this more
historical line, or, better to say, loop, we believe that it is
rather worth it to study GPDs from the AdS/QCD point of view.
Certainly, it is not expected that there is an exact answer for
(scheme dependent) partonic quantities; however, one might perhaps find  guidelines for
the parameterization of  GPDs with respect to their skewness  dependence.

\subsection*{Acknowledgements}

\noindent
D.M.~is grateful to the Rudjer Bo\v{s}kovi\'{c} Institute and the
Department of Physics at the University of Zagreb for an invitation
and the warm hospitality during his stay, during which this work has been
completed. He likes to thank his colleagues from Bochum for
numerous fruitful discussions, in particular, K.~Goeke,
P.~Pobylitsa, M.~Polyakov, and P.~Schweitzer. For discussion on
different aspects in this paper we would like to thank I.V.~Anikin,
S.~Brodsky, M.~Diehl, L.~Favart, M.~Furi\'{c}, M.~Guidal, V.~Guzey,
Ph.~H\"{a}gler, P.~Kroll, J.~Noritzsch, A.~V.~Radyushkin,
A.~Sch\"afer, L.~Schoeffel, O.~V.~Teryaev and M.~Vanderhaeghen.
K.K.~and D.M.~would also like to thank B.~Pire and F.~Sebastian for
invitation to the Journ\'ees GDR Nucl\'eon meeting ``Extracting
GPDs'', where this work has been presented in a preliminary form.
K.K.~would like to thank the Institut f\"{u}r Theoretische Physik
II at Ruhr-Universit\"{a}t Bochum for a kind hospitality. This
work was supported in part by the BMBF (Federal Ministry for
Education and Research), contract FKZ 06 B0 103 and by the
Croatian Ministry of Science, Education and Sport under 
contracts no. 119-0982930-1016 and 098-0982930-2864.


\end{document}